\documentclass[UTF8,a4paper]{article}
\usepackage{amsmath}
\usepackage{graphicx}
\usepackage{subfigure}
\usepackage{epstopdf}
\usepackage{geometry}
\usepackage{ulem}
\usepackage{indentfirst}
\usepackage{amsfonts,amssymb,dsfont}
\usepackage{setspace}
\usepackage{threeparttable}
\usepackage{amsmath,mathtools,amsthm}
\usepackage{array}
\usepackage{extarrows}
\usepackage{upgreek}

\makeatletter
\renewcommand*\env@matrix[1][\arraystretch]{%
  \edef\arraystretch{#1}%
  \hskip -\arraycolsep
  \let\@ifnextchar\new@ifnextchar
  \array{*\c@MaxMatrixCols c}}
\makeatother
\begin{document}


\title{\bf Electronic properties of the Dirac and Weyl systems with first- and higher-order dispersion in non-Fermi-liquid picture}
\author{Chen-Huan Wu
\thanks{chenhuanwu1@gmail.com}
\\College of Physics and Electronic Engineering, Northwest Normal University, Lanzhou 730070, China}

\maketitle
\vspace{-30pt}
\begin{abstract}
\begin{large}

We investigate the non-Fermi-liquid behaviors of the 2D and 3D Dirac/Weyl systems with low-order and higher order dispersion.
The self-energy correction,
symmetry,
free energy,
optical conductivity,
density of states,
and spectral function
are studied.
We found that, for Dirac/Weyl systems with higher order dispersion,
the non-Fermi-liquid features remain even at finite chemical potential,
and they are distinct from the ones in Fermi-liquid picture and the conventional non-Fermi-liquid picture.
The power law dependence of the physical observables on the energy as well as the logarithmic
renormalizations due to the long-range Coulomb interaction are showed.
The Landau damping of the longitudinal excitations within random-phase-approximation (RPA) for the non-Fermi-liquid case
are also discussed.\\
\\
$ Keywords$: non-Fermi-liquid behavior;
Dirac semimetal;
multi-Weyl semimetal;
self-energy correction;
longitudinal excitations.\\

\end{large}

\end{abstract}
\begin{large}

\section{Introduction}

Different to the normal metals where the Fermi-liquid theory is valid and 
the long-range Coulomb interaction can be ignored
due to the static screening by the particle-hole excitations\cite{Li X},
for
the topological insulator or when near the quantum critical in modern condensate matter physics,
the non-Fermi-liquid theory is required.
In Fermi-liquid theory, the excitations near the Fermi surface 
(usually within the order parameter fluctuation gap) are Fermionic,
that results in the uniform spin susceptibility\cite{Vekhter I} in contrast to the one in topological insulator\cite{Shiranzaei M2},
and also leads to the linear-temperature-dependence of the electronic specific heat rather that the logarithmic
divergent one as found in heavy-fermion system as well as the superconductors.
While at quantum critical point,
the order paramater quantum fluctuation is a gapless boson mode,
and leads to the non-Fermi-liquid behavior by coupling to the gapless Fermionic excitations.
Such coupling is accompanied by a strong Landau damping,
and with the small quasiparticle residue and large effective mass\cite{wums21,Sagi Y,Christensen R S},
which can be obtained from the self-energy and dispersion.
As the strong fluctuation, e.g., in strongly correlated systems,
is presented, the fixed point start from Fermi liquid in renormalization group flow is destabilized.
In this letter,
we investigate the non-Fermi-liquid behaviors of the 2D Dirac system with first-order dispersion in continuum model.
The exchange-induced Fermionic and Bosonic 
self-energy correction as well as other observable quantities are calculated and also for these of the
 Dirac/Weyl systems with higher dispersion.
Considering the disorder effect origin from the impurities,
the polaron as a excited quasiparticle
are important when consider the many-body effect (many-electron effect),
and the disorder-induced self-energy\cite{ele}
describes the impurity-Fermions (for Fermionic polaron) or impurity-Bosons (for Bosonic polaron) interaction and
with the impurity dressed by the corresponding particle-hole excitations.
Besides,
since the spin rotation is missing 
in the Dirac $\delta$-type impurity field,
the spin structure is fixed,
and the spin of impurity and that of majority particles are usually opposite,
that provies the opportunity to form the Cooper pair and the strongly bound dimer.

In the usual Dirac or Weyl systems,
the proportion of the interacting electron keeps increasing as the energy is lowered down to zero along the cone-like dispersion,
and finally at the zero-energy point, the noninteracting electrons still exist.
At higher energy, the umklapp scattering can open up a local charge gap at the middle point of Brillouin zone edge, i.e., the $M$-point,
and leads to a symmetry-breaking instability to breaks the Landau-Fermi liquid picture.
This can be implemented by the magnetic-doping\cite{Honerkamp C,Jia X},
which is in constract to the non-magnetic doping\cite{wums21}.
For high non-magnetic doping\cite{Honerkamp C,Attaccalite C} (with strong short-range interaction),
the saddle points are lifted above the Fermi level, and the effect of umklapp scattering is being lessened,
thus in such nonadiabatic case, the interaction (short-range) strength increases and induces some kinds of pairing instabilities like the 
$d$-wave superconductivity, which will not breaks the Fermi-liquid picture.
In the mean time, the nonadiabaticity leads to lower Fermi velocity.
This is in contrast to the case of long-range electron-electron interaction which is nearly unscreened in the 
charge neutrality point of the undoped Dirac or Weyl systems\cite{Elias D C},
where the conventional Landau theory is inapplicable,
and the self-energy and Fermi velocity can be logarithmically enhanced\cite{Elias D C,Vafek O}.
We have also obtain\cite{wums21} that, the large gap in $M$-point (which becomes largest at zero doping\cite{Attaccalite C}) 
also supports the adiabatic feature of the transport of electrons
through the Dirac cone or Weyl cone,
which corresponds to the non-Fermi-liquid picture.
This is also consistent with the GW calculation results\cite{Attaccalite C}
that the gap in $M$-point increases when the doping level closes to zero (i.e., the adiabatic non-Fermi liquid regime).
It is shown that even at low temperature where the umklapp scattering can be ignored, the
non-Fermi-liquid with very high resistivity can still appears by aid of the strong electron-acoustic phonon coupling\cite{Wu F}
which also brings the superconducting instability.

The Fermi-liquid to non-Fermi-liquid transition also affects the plasmon dispersion.
In Fermi-liquid picture with finite doping,
the short-range Hubbard-like interaction has both the density and spin channels with the charge-density
susceptibility and spin-density susceptibility.
The induced spin plasmon frequency acts as 
$\omega_{p}\sim\sqrt{q}$ for $q\ll k_{F}$ and $\sim \alpha q$ for $q\gtrsim k_{F}$ ($q$ is the scattering wave vector),
which is widely seen in the ordinary 2D Fermi-liquid.
In this regime, the dimensionless effective fine structure constant 
(Wigner-Seitz radius\cite{Culcer D})
$r_{s}=e^{2}/\hbar v_{F}\epsilon$ is small
due to the strongly screened Coulomb potential (in the limit of $\mu\gg\omega$ and $q\ll k_{F}$),
and thus the method of RPA provides an accurate result.
That can also be seen from the relation $r_{s}\propto 1/\sqrt{n}$ in 2D system,
which shows that the Coulomb interaction decreases with the raising doping level.
In the mean time, the gap in $M$-point as well as the fermi velocity decreases\cite{Elias D C}, and the phenomenon of
   Pauli blocking emergents during the interband transitions.
Note that in this limit, 
the transverse spin fluctuation,
if exist, will much stronger than the charge density fluctuation (with longitudinal excitations),
for example, 
in Helical liquid\cite{Raghu S} or even the chiral liquid system\cite{Jafari S A,Jalali-Mola Z} like magnetic graphene edges\cite{Yazyev O V}.
While for the non-Fermi-liquid picture in the zero-doping limit,
the preserved long-range Coulomb interaction has only the charge channel,
and the accuracy of RPA decreases with the increasing $r_{s}$.
The Coulomb potential is then obtained by\cite{Raghu S} $U=2\pi\hbar v_{F}r_{s}/q\sim r_{s}$.
In this case, if we focus on the low-energy regime with small scattering momentum $q$
and ignore the electron-phonon interaction,
the RPA susceptibility still gives the plasmon mode which $\sim\sqrt{q}$\cite{Raghu S}
and the system is still adiabatic,
however,
if we takes the phonon umklapp scattering into account and considering the momenta outside the first Brillouin zone 
which with higher energy,
the nonadiabatic feature will becomes more obvious with the phonon frequency $\Omega$ larger than Fermi energy $E_{F}$,
and that results in a quasilinear polaronic plasmon mode which behaves as $\sim \beta q$ (here $q$ is the phonon scattering momentum
and $\beta\ll \alpha$).
This has been experimental observed\cite{Zhang S,Jia X,Shvonski A} in the topological insulators.
In the mean time, the phonons decreases the group velocity of electrons near the nodes,
and exhibit nonadiabatic feature\cite{wums21}, and the quasiparticle resideu $Z$ also becomes lower in these nodes.
It is interesting that, even in the absence of phonon umklapp scattering (at low-energy regime),
the strong electron-phonon coupling could still breaks the Fermi-liquid picture.
That is because 
the strong many-body interaction effects can drive the
2D metallic Fermi liquid ground state into gapped insulator phase 
as the strong correlation goes beyond the Fermi-liquid paradigm\cite{Wu F,Sagi Y}.
Similar phenomenon can also be realized by the topological quantum phase transition through the merger of Weyl
nodes with opposite chirality\cite{Ahn S,Liu W E,Han S E}.
Such a largely increased resistivity with the breakdown of Fermi-liquid picture
has been observed recently in the (AA-stacked) twisted bilayer graphene near the magic angle\cite{Cao Y,Wu F} (at metallic phase) and the cuprates.
Such strange metallic behavior is been ascribed to the strong electron-acoustic phonon interaction
accompanied with the flattening of conduction band and valence band near the cone.
With the increace of electron-phonon interaction, the slope of band is lowered and thus the Fermi velocity of electron also
decrease.
That also leads to the nonadiabatic effects
with the enhanced interaction strength, like the decrease of quasiparticle residue $Z$ and increase of effective mass (not the Dirac mass).
While at low-temperature limit,
similar effect could also be observed (even more obvious) in the Weyl system with high-order dispersion.
It is because the dispersion at Dirac or Weyl cone has a decisive effect on the long-range Coulomb interaction.

In this paper, we consider the case of low-$q$ (without the internode scattering), and ignore the electron-phonon coupling
as well as the possible induced short-range electron-electron attraction\cite{Sentef M A}.
The differences maked by the high order dispersion
can be showed by the topological nature as well as the property of spin-momentum locking\cite{Ahn S},
also, the electronic properties like the self-energy and optical conductivity also modified by it.
The density-of-states (DOS) in the low-energy limit will be largely increased when the dispersion order $m$ turns up,
and accompanied with the enhanced dissipation of energy.
This may be the reason of the vanishing of residue $Z$ in low-energy limit of the multi-Weyl semimetals with high-order dispersion,
which is impossible for the linear Dirac or Weyl dispersion\cite{Wang J R},
although the
 screening to electron-electron interaction 
is heavily suppressed in neutral point of the undoped intrinsic Dirac or Weyl systems
as the DOS goes to zero.
We note that for 1D Luttinger liquid,
the compressibility will not vanishes even in zero residue\cite{Nandkishore R}.
In Sec.2, we discuss the self-energy correction in 2D Dirac system.
In Sec.3, we discuss the disorder effect and free energy.
The expressions of self-energy, optical conductivity and DOS of the Dirac and Weyl systems
characterized by the high order dispersion are analytically derived in the Sec.5.


\section{Self-energy correction in 2D Dirac system}

The isotropic (2+1)D Dirac Fermions coupled to long-range Coulomb interaction can be described by the effective action
\begin{equation} 
\begin{aligned}
S=\int d\tau d^{2}r \{\psi^{\dag}[\partial_{\tau}-ig\phi+H_{0}(k)]\psi+\frac{1}{2}[(\partial_{x}\phi)^{2}+(\partial_{y}\phi)^{2}]\},
\end{aligned}
\end{equation}
where $\psi$ is the Fermion field and $\phi$ is the Bosonic field which describes the long-range instantaneous Coulomb interaction
and related to the order parameter,
and the Fermions couples to the Bosons through the coupling constant $g^{2}=2\pi e^{2}/\epsilon>0$
where $\epsilon$ is the background dielectric constant.
Here we focus on the long-range Coulomb interaction,
while for the doped case with strong fluctuation\cite{},
the strong attraction within the short-range pairs may induces the Mott insulator phase\cite{Adibi E,Jafari S A2} or the superconductivity order parameter at finite doping\cite{Li X}.
The $H_{0}(k)$ is the non-interacting Hamiltonian which for 2D linear Dirac system reads
\begin{equation} 
\begin{aligned}
H_{0}(k)=\hbar v_{F}(\eta k_{x}\tau_{x}+k_{y}\tau_{y})+Ds_{z}\tau_{z}-\mu,
\end{aligned}
\end{equation}
whose eigenvalues can be obtained by solving ${\rm det}(H-E)=0$ as
\begin{equation} 
\begin{aligned}
E_{k\pm}=&\pm\sqrt{\hbar^{2}v_{F}^{2}k^{2}+D^{2}+\mu^{2}-2\eta\hbar v_{F}k\mu{\rm cos\theta}}-\mu\\
\approx &
\pm\sqrt{\hbar^{2}v_{F}^{2}k^{2}+D^{2}}-\mu,
\end{aligned}
\end{equation}
where $\theta={\rm arctan}\frac{k_{y}}{k_{x}}$.
By defining the scattered momentum as $k'=k+q$,
the exchange-induced Fermion self-energy is given by
\begin{equation} 
\begin{aligned}
\Sigma(q)=-g^{2}\int\frac{d\Omega}{2\pi}\frac{d^{2}k}{(2\pi)^{2}}G_{0}(\Omega+i0,k')D_{0}(k),
\end{aligned}
\end{equation}
and it's independent of the Boson Matsubara frequency $\omega=2m\pi T$
due to the instantaneous approximation of the Coulomb interaction
in one-loop order.
The instantaneous Coulomb interaction is given by the scalar potential
which has the propagator 
\begin{equation} 
\begin{aligned}
\langle T\phi(t,{\bf r})\phi(t',{\bf r}')\rangle=&
-i\delta(t-t')\frac{1}{2}\int\frac{d^{2}k}{(2\pi)^{2}}e^{i{\bf k}\cdot({\bf r}-{\bf r}')}\\
=&\frac{\delta(t-t')}{4\pi(r-r')},
\end{aligned}
\end{equation}
and that leads to the nonrelativistic features with the broken Lorentz invariance,
as widely seen in the non-perturbative RG analysis\cite{González J}
(while in perturbative RG analysis the instantaneous approximation is sometimes unadopted due to the effect of vector potential\cite{Wang J R}).
$G_{0}(\Omega+i0,k')=\frac{1}{\Omega+i0-H_{0}}$ 
is the bare Green's function (Fermion propagator).
The infinitesimal quantity $i0$ (corresponds to the scattering rate or the
Fermionic damping rate) is important for the convergence of the integral
and its sign is the same as that of the frequency 
(here we assuming the positive frequency).
The Pauli exclusion principle also enforces $i0\rightarrow 0$ in the static limit.
$D_{0}(k)=1/k$ is the bare Boson propagator.
Thus for static case, we have for a single Fermion species
\begin{equation} 
\begin{aligned}
\Sigma(q)=&-\frac{g^{2}}{(2\pi)^{2}}\left[
{\rm Li}_{2}(\frac{-D+\hbar v_{F}\Lambda+\mu+\Omega+i0}{\hbar v_{F}\Lambda})-
{\rm Li}_{2}(-\frac{-D-\hbar v_{F}\Lambda+\mu+\Omega+i0}{\hbar v_{F}\Lambda})\right.
\\&\left.+
{\rm ln}(D-\hbar v_{F}\Lambda-\mu-\Omega-i0){\rm ln}(-\frac{-D+\mu+\Omega+i0}{\hbar v_{F}\Lambda})\right.
\\&\left.-
{\rm ln}(D+\hbar v_{F}\Lambda-\mu-\Omega-i0){\rm ln}(\frac{-D+\mu+\Omega+i0}{\hbar v_{F}\Lambda})-
({\rm ln}(-\Lambda)-{\rm ln}\Lambda){\rm ln}(-D+\mu+\Omega+i0)\right]\bigg|_{\Omega}.
\end{aligned}
\end{equation}
where ${\rm Li}_{2}(z)=\sum^{\infty}_{k=1}\frac{z^{k}}{k^{2}}$ is the dilogarithm function.
Unlike the momentum shell integration, we only applied the ultraviolet cutoff here
to deal with the non-Fermi liquid with nearly zero gap (and chemical potential).
Ultraviolet cutoff during the calculation is important to prevent the divergence of integral.
At higher temperature, the repulsive Coulomb interaction is competes with the
attractive electron-phonon coupling.
The unscreened on-site Coulomb repulsion averts the double occupation of the lattice sites and thus closing the gap, while the 
electron-phonon coupling is opposite.
The lowest-order contribution to the exchange-induced self-energy reads
\begin{equation} 
\begin{aligned}
\Sigma(k,\omega)=T\sum_{\Omega}\int\frac{d^{2}q}{(2\pi)^{2}}U V(q,\Omega)G(k-q,\omega-\Omega),
\end{aligned}
\end{equation}
where $U$ is the Coulomb repulsion potential and 
\begin{equation} 
\begin{aligned}
V(q,\omega)=-\int\frac{d^{2}q}{(2\pi)^{2}}G(k,\Omega)G(k',\Omega'),
\end{aligned}
\end{equation}
is the fluctuation exchange potential.
Here we approximate the irreducible vertex function to the on-site Hubbard interaction,
and the resulting exchange self-energy is obviously beyond the instantaneous approximation

while for the attractive phonon-mediated interaction,
similarly, the self-energy reads
\begin{equation} 
\begin{aligned}
\Sigma(k,\omega)=T\sum_{\Omega}\int\frac{d^{2}q}{(2\pi)^{2}}U_{e-ph} P(q,\Omega)G(k-q,\omega-\Omega),
\end{aligned}
\end{equation}
with the phonon propagator (in lowest-order)\cite{Hague J P}
\begin{equation} 
\begin{aligned}
P(q,\Omega)=\frac{\Omega_{ph}^{2}}{\Omega_{ph}^{2}+\Omega^{2}},
\end{aligned}
\end{equation}
where $U_{e-ph}$ is the electron-phonon coupling parameter and $\Omega_{ph}$ is the phonon frequency.
For strong enough on-site attractive Hubbard interaction,
the charge-density-wave (CDW) phase or the gapless semimetal phase will be unstable to the s-wave superconducting phase,
and thus the symmetry described by $\langle\psi_{+}|\sigma_{x/y}|\psi_{-}\rangle=0$ is broken
($\pm$ refers to the up and down spin respectively) which don't consider the orbital degree of freedom.
For 2D Dirac semimetal,
due to the absence of the impurity scattering in the Dirac point with zero density of states,
the short-range interaction is weak and insufficient to destabilize the Dirac Fermions.
For superconducting phase without the Coulomb repulsion (instantaneous Coulomb interaction) and the disorder,
the Lorentz invariance is possible with isotropic Fermion and Boson velocities (i.e., in case of the supersymmetry
which interchanges bosons and fermions\cite{Grover T}),
which can be realized at low-energy by a metallic (polarizable) superstrate.
Through the minimum model in Eq.(2),
the time-reversal symmetry can be shown as
\begin{equation} 
\begin{aligned}
\Theta H_{0}(k)\Theta^{-1}=\tau_{y}H_{0}^{*}(k)\tau_{y}=\hbar v_{F}(-\eta k_{x}\tau_{x}-k_{y}\tau_{y})+D\tau_{z}\sigma_{z}-\mu=H_{0}(-k),
\end{aligned}
\end{equation}
where $\Theta=i\tau_{y}K$ is the time-reversal operator.
While for the 2D lattice model
\begin{equation} 
\begin{aligned}
H_{0}^{l}(k)=\sum_{k}c_{k}^{\dag}[t{\rm sin}k_{x}\tau_{y}+d_{z}(k)\tau_{z}]c_{k}-\sum_{k}c_{k}^{\dag}\mu c_{k},
\end{aligned}
\end{equation}
where $d_{z}(k)$ is the momentum-dependent gap function.
For this lattice model,
the particle-hole symmetry at half-filling can be revealed by
\begin{equation} 
\begin{aligned}
\Xi H_{0}^{l}(k)\Xi^{-1}=\tau_{y}H_{0}^{l*}(k)\tau_{y}=\sum_{k}c_{k}^{\dag}[-t{\rm sin}k_{x}\tau_{y}-d_{z}(k)\tau_{z}]c_{k}=H_{0}^{l}(-k),
\end{aligned}
\end{equation}
where $\Xi=\tau_{y}K$ is the particle-hole operator.
Topologically, $\Theta^{2}=0,\pm 1$ corresponds to the time-reversal symmetry and 
$\Xi^{2}=0,\pm 1$ corresponds to the particle-hole symmetry.
Although the time-reversal symmetry and the inversion symmetry are broken in the presence of gap function 
or by the charge-density-wave (CDW) order formed by polarized electrons,
the symmetry described by the product of $\Theta$ and the in-plane mirror reflection operator $M_{x}$ could be preserved\cite{Jian S K},
i.e., $\Theta M_{x}$ which protect the semimetal nature against the weak short-range interaction.
The weak short-range interaction can't be taken into account the RG analysis,
while the frequency-dependent self-energy in a non-Fermi liquid system 
is proportional to the anomalous dimension and the RG parameter (the logarithmic term).
The anomalous dimension also implies the missing of the pole struture of the Green's function,
which correspond to the electron addition and removal energies in the noninteracting case\cite{Onida G}.

In one-loop order,
the bare Boson propagator (phonon or the photon) is modified as $D(\Omega,k)=(k^{2}-\Sigma_{b}(\Omega,k))^{-1}$
where $\Sigma_{b}(\omega,k)$ is the Boson self-energy in density-density correlation form:
\begin{equation} 
\begin{aligned}
\Sigma_{b}(\omega,q)=&g_{s}g_{v}Tg^{2}\int\frac{d\Omega}{2\pi}\frac{d^{2}k}{(2\pi)^{2}}
{\rm Tr}[\sigma_{0}G_{0}(\Omega',k')\sigma_{0}G_{0}(\Omega,k)]\\
=&g_{s}g_{v}Tg^{2}\int\frac{d\Omega}{2\pi}\frac{d^{2}k}{(2\pi)^{2}}
4\frac{(-\Omega'-2i0-\mu)(-\Omega-i0-\mu)+kk'+D^{2}}{[(-\Omega'-2i0-\mu)^{2}-(k'^{2}+D^{2})][(-\Omega-i0-\mu)^{2}-(k^{2}+D^{2})]}\\
=&g_{s}g_{v}g^{2}\frac{1}{2\pi}\int\frac{d^{2}k}{(2\pi)^{2}}\frac{1}{4}
\sum_{ss'}(1+ss'\frac{kk'+D^{2}}{\varepsilon_{k}\varepsilon_{k'}})\frac{N_{F}(s\varepsilon_{k})-N_{F}(s'\varepsilon_{k'})}
{s\varepsilon_{k}-s'\varepsilon_{k'}+\omega+i0},
\end{aligned}
\end{equation}
where we use the discrete values of the frequency since otherwise the above integral becomes zero,
and the formula 
\begin{equation} 
\begin{aligned}
T\sum_{\Omega}
\frac{1}{-\Omega-i0-\mu+\varepsilon_{k}}
\frac{1}{-\Omega'-2i0-\mu+\varepsilon_{k'}}
=\frac{N_{F}(\varepsilon_{k})-N_{F}(\varepsilon_{k'})}{\varepsilon_{k}-\varepsilon_{k'}+\omega+i0}
\end{aligned}
\end{equation}
 is used.
Here $\Omega'=\Omega+\omega$
and $g_{s}g_{v}=2$ denotes the Fermion species (spin and valley degrees of freedom).
$N_{F}(x)=(1+e^{x/T})^{-1}$ is the Fermi-distribution function.
The above expression also implies that the Boson self-energy is related to the equation of motion for the bare Green's function.
Through the Ward identity $\gamma=\partial_{\omega}\Sigma$ which is independent of both the external frequency and the scattering wavevector,
the vertex function can be obtained as
\begin{equation} 
\begin{aligned}
\gamma=&-\frac{g^{2}}{2}\int\frac{d\Omega}{2\pi}\frac{d^{2}q}{(2\pi)^{2}}
{\rm Tr}[\sigma_{0}G_{0}(\Omega,q)\sigma_{0}G_{0}(\Omega,q)]\frac{1}{q^{2}-\Sigma_{b}(\Omega,q)}.
\end{aligned}
\end{equation}
To simplify the calculation, we restrict ourselves to the gapless case,
then the Boson self-energy becmoes
\begin{equation} 
\begin{aligned}
\Sigma_{b}(\Omega,q)=&g_{s}g_{v}g^{2}\frac{1}{(2\pi)^{2}}\int kdk\frac{1}{4}
(1-\frac{kk'}{\varepsilon_{k}\varepsilon_{k'}})\frac{1}
{\varepsilon_{k}+\varepsilon_{k'}-\Omega-i0}\\
=&g_{s}g_{v}g^{2}\frac{1}{(2\pi)^{2}}\frac{\hbar^{2}v^{2}-1}{16\hbar^{4}v^{4}}
\left\{
[(-\hbar v_{F}q+\Omega+i0){\rm ln}(-\hbar v_{F}(2\Lambda+q)+\Omega+i0)+2\hbar v_{F}\Lambda]\right.\\
&\left.-[(-\hbar v_{F}q+\Omega+i0){\rm ln}(-\hbar v_{F}q+\Omega+i0)]\right\},
\end{aligned}
\end{equation}
where $ss'=-1$ here due to the dominating interband transition.
Then the vertex function at half-filling is
\begin{equation} 
\begin{aligned}
\gamma=&-\frac{g^{2}}{2(2\pi)^{2}}
\int d\Omega dq
\frac{q}{-\hbar v_{F}q+\Omega+i0}\\
&[q^{2}-2\alpha\hbar v_{F}\Lambda+\alpha(-\hbar v_{F}q+\Omega+i0){\rm ln}(-\hbar v_{F}q+\Omega+i0)\\
&-\alpha(-\hbar v_{F}q+\Omega+i0){\rm ln}(-\hbar v_{F}(q+2\Lambda)+\Omega+i0)]^{-1}.
\end{aligned}
\end{equation}

\section{Disorder effect and free-energy in grand-canonical ensemble}

For the disordered system, the Fermion self-energy in lowest-order approximation reads
$\Sigma^{LO}=n_{i}V_{k,k'}/\hbar$ where $n_{i}$ is the impurity concentration and $V_{k,k'}$ is the scattering potential.
For localized potential, the disorder-induced self-energy could be momentum-independent
due to the rotational invariance.
In low-order perturbation theory, the self-energy matrix contains a gap function in the diagonal element,
while the non-diagonal elements are missing.

The random-phase approximation (RPA) results is valid only in the long-wavelength limit as well as
the low-energy limit for large flavor number analysis,
in which case the Eliashberg theory as well as the Migdal's theorem are valid.
In this case the Boson propagator (as well as the Boson-frequency-related spin susceptibility) 
is overdamped due to the small Boson velocity
and small external Boson momentum (compared to the Fermionic ones) with the Landau damping.
The above results are correct for the low-energy Fermions excitations (within the band gap) 
for RPA which with chemical potential much larger than $k_{B}T$.
Inversely,
the non-Fermi-liquid feature emerges for the case of $\mu<k_{B}T$.
The strong screeing effect by polarized Fermions to the disorder also provides the possibility to recover the Fermi liquid 
within the spectrum gap of the order parameter fluctuations 
(of the order of $D^{2}/W$ where $W$ is the Fermion bandwidth) which with coherent Bosonic excitations except when the disorder-induced
linewidth\cite{Motome Y} is larger than the excitation energy.
In the mean time, the excitation gap gives rise to the dissipation effect
which is related to the free energy and the conductivity.
Further, the response function is nonzero even at $q=0$
for the Bosonic frequency in the range $\omega<v_{F}q<2D<2\mu$.
While for the Bosonic frequency $\Omega$ larger than $v_{F}q$,
the transverse spin excitation (still within the band gap) is the Goldstone spin wave and thus it's gapless in the long-wavelength limit,
in contrast to the longitudinal excitations which is gapped even at $q=0$.

Unlike the weak short-range interaction,
the electron-electron interaction mediated by the gapless Bosonic mode is long ranged at the quantum critical point and
with the gapless critical fluctuation of the order parameter (about the Bosonic excitations)
which can be described by the Ginzburg-Landau function.
The Ginzburg-Landau function (the free energy) here describing the order parameter fluctuation does not contains the term 
$\phi^{*}\partial_{\tau}\phi$ due to the particle-hole symmetry
as state above, where $\phi$ is the two component complex amplitude.
For bipartite system, the particle-hole symmetry suggests the exitence of the zero-energy modes which satisfy 
$\phi_{A}=\pm i\phi_{B}$.
In the presence of particle-hole symmetry the ac Hall conductivity vanishes 
while the dc conductivity is preserved\cite{Van Otterlo A}.
At finite temperature, the free energy can be obtained by the following partition function base on the Fermion propagator as
\begin{equation} 
\begin{aligned}
\mathcal{Z}={\rm det}\left[\frac{1}{G_{0}(k,\Omega)T}\right]=\prod_{\varepsilon}\left(\frac{\Omega+i0-H_{0}}{T}\right)^{2}_{\varepsilon},
\end{aligned}
\end{equation}
then the free energy density reads
\begin{equation} 
\begin{aligned}
F=&-T\int \frac{d^{2}k}{(2\pi)^{2}}\sum_{i\Omega=0}^{n}{\rm ln}\mathcal{Z}
=&-T\lim_{n\rightarrow \infty}\int \frac{d^{2}k}{(2\pi)^{2}}
{\rm ln}[\frac{H_{0}^{2}}{T^{2n+2}}(\frac{\Gamma(1-H_{0}+n)}{\Gamma(1-H_{0})})^{2}].
\end{aligned}
\end{equation}
The above integral can be analytically solved as
\begin{equation} 
\begin{aligned}
F=&-T\lim_{n\rightarrow \infty}\frac{1}{2\pi 2\hbar^{2}v_{F}^{2}}
[-2D^{2}{\rm ln}(H_{0})-2\hbar^{2}v_{F}^{2}k^{2}{\rm ln}\Gamma(-H_{0}+n+1)
+2\hbar^{2}v_{F}^{2}k^{2}{\rm ln}\Gamma(-H_{0}+1)\\
&+\hbar^{2}v_{F}^{2}k^{2}{\rm ln}(\frac{H_{0}^{2}}{T^{2n+2}}(\frac{\Gamma(-H_{0}+1+n)}{\Gamma(-H_{0}+1)})^{2})
-4\psi^{(-3)}(-H_{0}+n+1)+4\psi^{(-3)}(-H_{0}+1)\\
&-4\hbar v_{F}k\psi^{(-2)}(-H_{0}+n+1)+4\hbar v_{F}k\psi^{(-2)}(-H_{0}+1)
-2\mu^{2}{\rm ln}(H_{0})
+4D\mu{\rm ln}(H_{0})\\
&+2D\hbar v_{F}k-\hbar^{2}v_{F}^{2}k^{2}-2\hbar v_{F}k\mu]\bigg|_{k},
\end{aligned}
\end{equation}
and for semimetal at half-filling it reduces to
\begin{equation} 
\begin{aligned}
F=&-T\lim_{n\rightarrow \infty}\frac{1}{2\pi 2\hbar^{2}v_{F}^{2}}
[-2\hbar^{2}v_{F}^{2}k^{2}{\rm ln}\Gamma(-\hbar v_{F}k+n+1)+\hbar^{2}v_{F}^{2}k^{2}{\rm ln}
(\frac{\hbar^{2}v_{F}^{2}k^{2}}{T^{2n+2}}(\frac{\Gamma(1-\hbar v_{F}k+n)}{\Gamma(1-\hbar v_{F}k)})^{2})\\
&-\hbar^{2}v_{F}^{2}k^{2}
-4\hbar v_{F}k\psi^{(-2)}(n-\hbar v_{F}k+1)
-4\psi^{(-3)}(-\hbar v_{F}k+n+1)\\
&+4\hbar v_{F}k\psi^{(-2)}(-\hbar v_{F}k+1)
+4\psi^{(-3)}(-\hbar v_{F}k+1)
+k^{2}{\rm ln}\Gamma(1-\hbar v_{F}k)]\bigg|_{k},
\end{aligned}
\end{equation}
where ${\rm ln}\Gamma$ denotes the logarithm of the Gamma function
and $\psi^{(n)}$ is the $n$th derivative of the digamma function.
By using the aproximational relation\cite{Han S E}
\begin{equation} 
\begin{aligned}
{\rm ln}(\frac{\omega^{2}+H_{0}^{2}}{T^{2}})\approx \frac{H_{0}}{T}+2{\rm ln}(1+e^{-H_{0}/T}),
\end{aligned}
\end{equation}
then the free energy density can be rewritten as
\begin{equation} 
\begin{aligned}
F=-T\frac{1}{2\pi}
&[\frac{2T^{2}{\rm Li}_{3}(-e^{\hbar v_{F}k/T})}{\hbar^{2}v_{F}}
+\frac{1}{3}k^{2}(\frac{2\hbar v_{F}k}{T}+3{\rm ln}N_{F}(-\hbar v_{F}k)-3{\rm ln}N_{F}(\hbar v_{F}k))\\
&-\frac{2kT{\rm Li}_{2}(-e^{\hbar v_{F}k/T})}{\hbar v_{F}}]
\bigg|_{k},
\end{aligned}
\end{equation}
where ${\rm Li}_{n}(x)$ is the polylogarithm function.
Consider the many-body effect to the 2D Dirac system,
the perturbations can be taken into account in the grand-canonical ensemble,
where we rewrite the tight-binding model Hamiltonian as
\begin{equation} 
\begin{aligned}
H=&H_{0}+H_{E}+H_{D}\\
=&\sum_{ij}(tc^{\dag}_{i}c_{j})+\sum_{ij}\frac{1}{2}U_{ij}n_{\uparrow}n_{\downarrow}\\
&+V_{0}\sum_{i}c^{\dag}_{i}c_{i}\delta(r-r_{i}),
\end{aligned}
\end{equation}
where $ij$ is the nearest neighbor sites and $t$ is the nearest neighbor hopping.
$n=c^{\dag}_{i}c_{j}$.
$U_{ij}$ is the Coulomb interaction strength in second term which is the Coulomb exchange interaction-related term (bilinear).
$V_{0}$ is the impurity scattering potential (magnetic or nonmagnetic) in the third term which is the disorder-related term.
The creation and annihilate operators are all the particle one here.
Then the free energy density (grand potential) is still $F=-T{\rm ln}Z$
but with the partion function in interacting case as
\begin{equation} 
\begin{aligned}
Z(i\Omega,U_{ij},V_{0})=&{\rm Tr}{\rm exp}[\frac{-H+N\mu}{T}]\\
=&\int D\psi D\psi^{*}e^{S},
\end{aligned}
\end{equation}
where the path integral runs over the Grassman variables, $N$ is the particle number operator,
and
with the action $S$ reads
\begin{equation} 
\begin{aligned}
S(i\Omega,U_{ij},V_{0})=&\sum_{i\Omega}\psi(i\Omega)[(i\Omega+\mu)\delta_{ij}-H_{0}-H_{D}]\psi^{*}(i\Omega)
-\int^{1/T}_{0}d\tau H_{E}(\tau),
\end{aligned}
\end{equation}
where $\psi(i\Omega)$ is the real Grassmann variable,
and the first term is always positive summing over the Fermionic frequency.
The perturbaed Green's function which satisfies the Dyson relation
$G=G_{0}+G\Sigma G_{0}$ (or $\Sigma=G_{0}^{-1}-G^{-1}$),
can be obtained by the ratio of the partion functions as
\begin{equation} 
\begin{aligned}
G(i\Omega,U_{ij},V_{0})=&\frac{-Z'(i\Omega,U_{ij},V_{0})}{Z(i\Omega,U_{ij},V_{0})}\\
=&\frac{-\int D\psi D\psi^{*}e^{S}\psi(i\Omega)\psi^{*}(i\Omega)}{\int D\psi D\psi^{*}e^{S}},
\end{aligned}
\end{equation}
The case of $G(i\Omega,U_{ij},V_{0})\neq 1$ clearly indicates broken of the supersymmetry\cite{Ziegler K}.
We consider the $\delta$-type impurity potential in above disorder-term,
which indicates the Born approximation.
In such case the spin structure is fixed as also been observed in the surface state of the 3D topological insulators,
thus the spin rotation is missing 
which can be observed in the nodal-line semimetals,
and the spin current operator becomes zero in the helicity basis.
The Born approximation guarantees the sign-invariance of the momentum before and after the scattering,
and the reversed scattering amplitude has the same value with the origin one\cite{Khaetskii A},
In grand canonical ensemble, the spin current vanishes in the thermodynamical limit (and thus with infinite $\mu$)
due to the vanishing spin density even when beyong the Born approximation.
Beyong the $\delta$-type impurity field, the extrinsic spin current emergents and 
the scattering of both the impurity and the majority particle (with opposite spins) create the Fermionic polaron,
and the optical Hall conductivity (in fact this is the only case where the transverse conductivity equals to the Hall conductivity)
of the polaron determines the current in direction orthogonal to the external force\cite{Camacho-Guardian A},
is related to the current operator by 
$\sigma_{xy}=J_{x}/(-\nabla_{y}V)$ in linear response theory where $V$ is the external potential.
The current operator here is much smaller than the one in QED just like the group velocity operator which is much smaller than the speed of
light.
While for the Bosonic polaron for the system immersed in the Bose-Einstein condensates,
the interaction is stronger than the Fermionic one
due to the higher compresibility of the BEC compared to the Fermionic media.
The Bosonic polaron is formed by the Fermionic impurity which dressed by the majority Bosonic excitations.

\section{Optical conductivity}

The destoryed Fermi-liquid behavior can be observed by the singular Bosonic susceptibility at the nesting wavevector,
and it can't be found even at low-energy limit (far away from the quantum critical point)
when the ultraviolet cutoff applied is infinite during the calculation,
as could be found in the many-electron system, 
for example, when the Fermions coupled to the 1D Ising variable\cite{Yin S} or the fluctuation transverse gauge field\cite{Mross D F}
or the longitudinal Bosonic excitation\cite{Vekhter I}.
The non-Fermi-liquid phenomenons are widely observed in the heavy-Fermi system and the cuprate materials\cite{Senthil T,VV},
including the logarithmic divergent specific heat 
which related to the free energy by
$C_{V}=-T\frac{\partial^{2}F}{\partial T^{2}}$.
In the presence of monochromatic light, the nondiagonal part of the optical conductivity can be
obtained by summing over the eigenvalues:
\begin{equation} 
\begin{aligned}
\sigma_{xy}(\omega)&=i\hbar e^{2}\int\frac{d^{2}k}{(2\pi)^{2}}\sum_{\varepsilon,\varepsilon'}\frac{[N_{F}(\varepsilon)-N_{F}(\varepsilon')]
\langle\varepsilon |v_{x}|\varepsilon'\rangle \langle \varepsilon'|v_{y}|\varepsilon \rangle}{(\varepsilon-\varepsilon')(\varepsilon-\varepsilon'+\omega+i0)}\\
=&i\hbar e^{2}\int\frac{d^{2}k}{(2\pi)^{2}}\sum_{\varepsilon,\varepsilon'}\frac{[N_{F}(\varepsilon)-N_{F}(\varepsilon')]
\frac{iv_{F}^{2}}{4}[(1-\frac{D}{\varepsilon})(1+\frac{D}{\varepsilon'})-(1-\frac{D}{\varepsilon'})(1+\frac{D}{\varepsilon})](1-\delta_{ss'})}
{(\varepsilon-\varepsilon')(\varepsilon-\varepsilon'+\omega+i0)},
\end{aligned}
\end{equation}
where only the retarded Green's function is used in contrast to the Streda one\cite{Streda P}.
Here the identify 
\begin{equation} 
\begin{aligned}
N_{F}(\varepsilon)(1-N_{F}(\varepsilon'))(1-e^{\varepsilon/T})=N_{F}(\varepsilon)-N_{F}(\varepsilon').
\end{aligned}
\end{equation}
is used.
The velocity matrix elements are
\begin{equation} 
\begin{aligned}
\langle \varepsilon |v_{x}|\varepsilon'\rangle=&\frac{v_{F}}{2}[s\sqrt{1+\frac{D}{\varepsilon}}\sqrt{1-\frac{D}{\varepsilon'}}
+s'\sqrt{1-\frac{D}{\varepsilon}}\sqrt{1+\frac{D}{\varepsilon'}}](1-\delta_{ss'}),\\
\langle \varepsilon'|v_{y}|\varepsilon \rangle=&\frac{iv_{F}}{2}[s'\sqrt{1+\frac{D}{\varepsilon'}}\sqrt{1-\frac{D}{\varepsilon}}
-s\sqrt{1-\frac{D}{\varepsilon}}\sqrt{1+\frac{D}{\varepsilon'}}](1-\delta_{ss'}),
\end{aligned}
\end{equation}
where the electron/hole indices have $ss'=-1$ during the optical transition due to the Pauli exclusion principle,
while the spin indices $\sigma_{z}$ before and after transition are invariant when (both the intrinsic and extrinsic) 
Rashba-coupling are negligible, otherwise the spin index changes since it's nomore the good quantum number.
In case that the Fermi level lies within the band gap, the diagonal elements of the conductivity are zero
(and thus implies the $C_{4}$ symmetry of the system since $\sigma_{xx}=\sigma_{yy}$),
while the non-diagonal elements becomes independent of the Dirac-mass due to the vanishing classical term\cite{Streda P}.
Note that the result of $\sigma_{xx}=\sigma_{yy}$ as well as the $\sigma_{xy}=-\sigma_{yx}$ also appear in the optical limit 
with $q\rightarrow 0$ (also called the local limit)
Taking into account the effect of chemical potential,
then the Hall conductivity is composed of two parts:
\begin{equation} 
\begin{aligned}
\sigma_{xy}(\omega)=&i\hbar e^{2}\int\frac{d^{2}k}{(2\pi)^{2}}
\Theta(\Lambda-\varepsilon)[\frac{\langle\varepsilon |v_{x}|\varepsilon'\rangle \langle \varepsilon'|v_{y}|\varepsilon \rangle}
{(\varepsilon-\varepsilon')(\varepsilon-\varepsilon'+\omega+i0)}
+\frac{\langle\varepsilon' |v_{x}|\varepsilon\rangle \langle \varepsilon|v_{y}|\varepsilon' \rangle}
{(\varepsilon-\varepsilon')(-\varepsilon+\varepsilon'+\omega+i0)}],\\
\sigma_{xy}(\omega,\mu)=&i\hbar e^{2}\int\frac{d^{2}k}{(2\pi)^{2}}
\Theta(\mu-\varepsilon)\{
[\frac{\langle\varepsilon |v_{x}|\varepsilon\rangle \langle \varepsilon|v_{y}|\varepsilon \rangle}
{(\varepsilon+\varepsilon')(-\varepsilon-\varepsilon'+\omega+i0)}
+\frac{\langle\varepsilon |v_{x}|\varepsilon\rangle \langle \varepsilon|v_{y}|\varepsilon \rangle}
{(\varepsilon+\varepsilon')(\varepsilon+\varepsilon'+\omega+i0)}]\\
&-
[\frac{\langle\varepsilon |v_{x}|\varepsilon'\rangle \langle \varepsilon'|v_{y}|\varepsilon \rangle}
{(\varepsilon-\varepsilon')(\varepsilon-\varepsilon'+\omega+i0)}
+\frac{\langle\varepsilon' |v_{x}|\varepsilon\rangle \langle \varepsilon|v_{y}|\varepsilon' \rangle}
{(\varepsilon-\varepsilon')(-\varepsilon+\varepsilon'+\omega+i0)}]\},
\end{aligned}
\end{equation}
where the first part corresponds to the case that the chemical potential is small than the Dirac-mass while the second
part is opposites.
While the longitidinal optical conductivity reads
\begin{equation} 
\begin{aligned}
\sigma_{xx}(\omega)
=&\sigma_{xx}^{inter}(\omega)
=&-e^{2}\pi\hbar\int\frac{d^{2}k}{(2\pi)^{2}}
\frac{|\langle \varepsilon|v_{x}|\varepsilon'\rangle|^{2}}{\varepsilon-\varepsilon'}\delta(\omega+i0+\varepsilon-\varepsilon'),
\end{aligned}
\end{equation}
where the intraband part vanishes unless at finite temperature and at infrared limit (nearly zero photon energy).

Different to the hopping-current-related conductivity,
the dissipation-current-related conductivity remains finite in static limit and proportional to $i/\omega+\pi\delta(\omega)$\cite{Van Otterlo A}
where $\delta$ is the Dirac-$\delta$ function here,
but this part of the conductivity is negligible when under a magnetic field or at low-temperature.
For the case of large band gap, the frequency about the optical tansition during the intraband process is much larger than the interband one
as shown in the WSe$_{2}$\cite{Tahir M}.

\section{Observable quantities in Dirac/Weyl systems with higher-order dispersion}

\subsection{2D Dirac system}

Next we discuss the 2D topological insulator (TI) with higher-order dispersion
and small momentum-dependent mass term,
whose Hamiltonian reads 
\begin{equation} 
\begin{aligned}
H_{0}^{m}=\frac{k_{0}^{2-m}a}{2}\hbar v_{F} |{\bf k}|^{m}(\hat{k}\cdot\sigma)+(c_{1}\sigma_{z}+c_{2}|{\bf k}|^{m}\sigma_{z}\tau_{z})-\mu,
\end{aligned}
\end{equation}
where we assume the Dirac-mass is momentum-dependent and controlled by the material-related constant $c_{1}$ and $c_{2}$,
the $k_{0}$ is another material-related constant in unit of momentum\cite{Ahn S} and $a$ is the lattice spacing\cite{GEO},
e.g., it equals to $\sqrt{3}/2$ times of the lattice constant in the graphene-like hexagonal lattice system.
In the following we denote $\xi=\frac{k_{0}^{2-m}a}{2}\hbar v_{F}$.
The term $(\hat{k}\cdot\sigma)$ here only appears in the chiral systems with spin-momentum locking,
while for the non-chiral systems, it's usually replaced by the spin operator $\sigma_{z}$
and the interband transition also vanishes in such case.
The (in-plane) momentum-dependent mass term $(c_{1}+c_{2}|{\bf k}|^{m})$ 
here is similar to the effect of next-nearest-neighbor (intrinsic) Rashba coupling.
Here we note that, we try to present a discussion for in the 2D Dirac system extend to the generic order $m$
which also related to the Chern number in gapless case,
and it's not just applicable to the 2D TI, 
but also to the multilayer TI which with a single 2D Dirac node per surface Brillouin zone\cite{Burkov A A,Shiranzaei M}.
The order $m$ controls the in-plane band dispersion,
for example,
$m=1$ for the (topologically protected; which 
not exists currently in 3D real space\cite{Park S}) 
linear Dirac dispersion, $m=2$ for the quadratic dispersion, $m=3,4$ for the trigonal warping system
as found in the monolayer MoS$_{2}$\cite{Scholz A}.
The above low-energy effective Hamiltonian of the high-order topological Dirac system can be rewritten 
in the tight-binding approximation as
\begin{equation} 
\begin{aligned}
H_{0}^{m}=\int\frac{d^{2}k}{(2\pi)^{2}}
\psi^{\dag}_{k\sigma_{z}}H_{k}\psi_{k\sigma_{z}}-\mu,
\end{aligned}
\end{equation}
where $\psi^{\dag}_{k\sigma}=(c^{\dag}_{k\sigma_{z}} d^{\dag}_{k\sigma_{z}})$ is the spinor field
with two creation operators correspond to the two sublattices degrees of freedom, and
\begin{equation} 
\begin{aligned}
H_{k}=\xi \begin{pmatrix}
c_{1}+c_{2}|{\bf k}|^{m} & (k_{x}-ik_{y})^{m} & 0 & 0\\
   (k_{x}+ik_{y})^{m}            & c_{1}-c_{2}|{\bf k}|^{m} & 0 & 0\\
 0 & 0 & c_{1}-c_{2}|{\bf k}|^{m} & (k_{x}-ik_{y})^{m} \\
 0 & 0 &  (k_{x}+ik_{y})^{m}            & c_{1}+c_{2}|{\bf k}|^{m}
\end{pmatrix}.
\end{aligned}
\end{equation}

In noninteracting case, the Bosonic propagator is overdamped with the gapless longitudinal excitations 
(or order parameter fluctuation) by the Landau damping,
it's also found that the Landau damping of the multi-Weyl semimetal is weaker than that of the marginal Fermi liquid\cite{Wang J R}
which is distinct from the normal non-Fermi-liquid states.
The dispersion of the multi-node Dirac system can be obtained by solving above Hamiltonian as
\begin{equation} 
\begin{aligned}
\varepsilon^{m}=&\pm \sqrt{c_{1}^{2}+2c_{1}c_{2}k^{m}+c_{2}^{2}k^{2m}+\mu^{2}+k^{2m}\xi^{2}-2\eta |{\bf k}|^{m}\mu\xi {\rm cos}\theta}\\
\approx &\pm \sqrt{c_{1}^{2}+2c_{1}c_{2}k^{m}+c_{2}^{2}|{\bf k}|^{2m}+|{\bf k}|^{2m}\xi^{2}}\\
=&\pm \sqrt{(c_{1}+c_{2}|{\bf k}|^{m})^{2}+|{\bf k}|^{2m}\xi^{2}},
\end{aligned}
\end{equation}
where we rewrite the term $\frac{k_{0}^{2-m}a}{2}\hbar v_{F}$ as the scale-dependent parameter $\xi$.

For the above multi-node dispersion,
the eigenvectors for conduction band and valence band are
\begin{equation} 
\begin{aligned}
|\Psi_{k}^{\lambda}\rangle=&\frac{e^{i{\bf k}\cdot {\bf r}}}{\sqrt{S}}\frac{1}{\sqrt{2}}
\begin{pmatrix}
{\rm cos}\frac{\alpha}{2}\\
-\lambda{\rm sin}\frac{\alpha}{2}e^{\lambda im\theta }
\end{pmatrix},\\
\end{aligned}
\end{equation}
where $\alpha={\rm arctan}\frac{\xi |{\bf k}|^{m}}{(c_{1}+c_{2}|{\bf k}|^{m})}$ and 
$\theta ={\rm arctan}\frac{k_{y}}{k_{x}}$
is the polar angle of ${\bf k}$.
$\lambda=\pm 1$ correspond to the electron and hole states, respectively.
Thus for electron and hole states, we have the following spinor parts
\begin{equation} 
\begin{aligned}
|\Psi_{k}^{+}\rangle_{s}=&\frac{1}{\sqrt{2}}
\begin{pmatrix}
{\rm cos}\frac{\alpha}{2}\\
-{\rm sin}\frac{\alpha}{2}e^{im\theta }
\end{pmatrix},\\
|\Psi_{k}^{-}\rangle_{s}=&\frac{1}{\sqrt{2}}
\begin{pmatrix}
{\rm sin}\frac{\alpha}{2}\\
{\rm cos}\frac{\alpha}{2}e^{im\theta }
\end{pmatrix}.\\
\end{aligned}
\end{equation}
Then the overlap factor (the form factor within the electronic susceptibility) for this chiral model reads
\begin{equation} 
\begin{aligned}
|\langle \Psi^{\lambda'}_{k+q}|\Psi^{\lambda}_{k}\rangle|^{2}=&
\left(\frac{e^{-i{\bf k'}\cdot{\bf r}+i{\bf k}\cdot{\bf r}}}{2S}\right)^{2}
\left[
{\rm cos}\frac{\alpha}{2}{\rm cos}\frac{\alpha'}{2}+\lambda\lambda'
{\rm sin}\frac{\alpha}{2}{\rm sin}\frac{\alpha'}{2}
e^{\lambda im(\theta -\theta')}
\right]^{2}\\
=&\frac{e^{2i({\bf k}\cdot{\bf r}-{\bf k}'\cdot{\bf r})}}{4S^{2}}
  \left[
{\rm cos}(\frac{\alpha}{2}-\frac{\alpha'}{2})(-1-\lambda\lambda'e^{i\lambda m(\theta-\theta')})+
{\rm cos}(\frac{\alpha}{2}+\frac{\alpha'}{2})(-1+\lambda\lambda'e^{i\lambda m(\theta-\theta')}   )
  \right]^{2},
\end{aligned}
\end{equation}
where ${\bf k}'={\bf k}+{\bf q}$, $\theta'={\rm arctan}\frac{k'_{y}}{k'_{x}}$,
and $\alpha'={\rm arctan}\frac{\xi |{\bf k}'|^{m}}{(c_{1}+c_{2}|{\bf k}'|^{m})}$.
The form factor here ensuring the electronic susceptibility only contributed by the overlap of electron-hole pair
(i.e., the scattering between conduction band and valence band).
It's obvious that we do not consider the effect of spin state here,
which means that we only consider the density-density correlation which is in longitudinal channel.
In another case, if the spin or pseudospin degrees of freedom are also been traced over,
the current tensors should be discussed in diagonal and non-diagonal cases\cite{Jalali-Mola Z,Jalali-Mola Z2,Jalali-Mola Z3}.
However,
if we restrict on the longitudinal channel with weak short-range interaction (in non-Fermi liquid picture),
we found that no matter for the graphene-like Dirac systems or the helical liquid systems
(with eigenvectors 
$\frac{e^{i{\bf k}\cdot {\bf r}}}{\sqrt{S}}\frac{1}{\sqrt{2}}
({\rm cos}\frac{\alpha}{2} ,-i\lambda{\rm sin}\frac{\alpha}{2}e^{i\lambda m\theta})^{T}$),
which couples with the momentum by the pseudospin and spin degrees of freedom respectively,
the overlap function is the same (i.e., Eq.(40)),
while for the massless case,
their overlap function is ${\rm cos}^{2}(\frac{m\theta'-m\theta}{2})$.

At half-filling, the Fermi level crosses the multi-band touching point and the Coulomb interaction remains long-ranged
as described by the Bosonic field,
due to the poor screening to the electron-electron Coulomb interaction.
The imaginary part of the exchange-induced self-energy is related to the quasiparticle relaxation time (lifetime),
while the real part is related to the interaction strength and the quaisparticle weight.
Beyond the instantaneous approximation induced by the scalar potential, 
the exchange-induced self-energy containing Bosonic frequency reads
(see Appendix.A for a further calculation and discussion)
\begin{equation} 
\begin{aligned}
\Sigma^{m}(\omega,k)=-g^{2}\int\frac{d\Omega}{2\pi}\frac{d^{2}k}{(2\pi)^{2}}G_{0}(\omega+\Omega,k')\frac{1}{q-\Sigma_{b}(\Omega,q)},
\end{aligned}
\end{equation}
where the last term of the above expression is the dressed Coulomb potential.
$\Sigma_{b}^{m}(\Omega,q)$ is the Bosonic self-energy (i.e., the dynamical polarization here)
which reads
\begin{equation} 
\begin{aligned}
\Sigma_{b}^{m}(\Omega,q)=&g_{s}g_{v}g^{2}\int\frac{d\Omega}{2\pi}\frac{d^{2}k}{(2\pi)^{2}}
{\rm Tr}[\sigma_{0}G^{m}_{0}(\Omega',k')\sigma_{0}G^{m}_{0}(\Omega,k)].
\end{aligned}
\end{equation}
then the multi-Dirac-node bare Green's function reads
\begin{equation} 
\begin{aligned}
G^{m}_{0}(\Omega,k)=&\frac{\Omega+i0+H_{0}^{m}}{(\Omega+i0)^{2}-H^{m2}_{0}}\\
=&\frac{\Omega+i0+\xi k^{m}(\hat{k}\cdot\sigma)+(c_{1}+c_{2}k^{m})\sigma_{z}\tau_{z}-\mu}
{(\Omega+i0)^{2}-(\xi k^{m}(\hat{k}\cdot\sigma)+(c_{1}+c_{2}k^{m})\sigma_{z}\tau_{z}-\mu)^{2}}.
\end{aligned}
\end{equation}
Base on this Green's function,
the dynamical polarization is available by the above expression but it's too verbose to express
which contains a hypergeometric function whose parameters are all related to the order $m$.
We then turn to a more concise expression at zero-temperature limit which reads
\begin{equation} 
\begin{aligned}
\Sigma_{b}^{m}(\Omega,q)&=\int\frac{d^{2}k}{(2\pi)^{2}}\sum_{ss'}
\frac{g_{s}g_{v}}{2}\frac{1+ss'{\rm cos}b}{\Omega+i0+s\varepsilon^{m}-s'\varepsilon^{m'}},\\
{\rm cos}b&=\frac{k^{m}+q{\rm cos}a}{\sqrt{k^{2}+q^{2}+2k^{m}q{\rm cos}a}}=1-\frac{q^{2}{\rm sin}^{2}a}{2k^{2m}}+O(q^{3}),\\
\varepsilon^{m}-\varepsilon^{m'}&=
-\frac{mqk^{m-1}(c_{1}c_{2}+c_{2}^{2}k^{m}+\xi^{2}k^{m})}{\sqrt{c_{1}^{2}+2c_{1}c_{2}k^{m}+c_{2}^{2}k^{2m}+\xi^{2}k^{2m}}}+O(q^{2}),
\end{aligned}
\end{equation}
where $b$ is the angle between $k$ and $k'$ and $a$ is the angle between $k$ and $q$.
The dynamical polarization can be devided into the intraband and interband parts:
\begin{equation} 
\begin{aligned}
\Sigma_{b}^{m}(\Omega,q)&=\Sigma_{b}^{m,intra}(\Omega,q)+\Sigma_{b}^{m,inter}(\Omega,q),\\
\Sigma_{b}^{m,intra}(\Omega,q)&=
\frac{g_{s}g_{v}}{2}\int^{2\pi}_{0}d\theta\int^{k_{F}}_{0}kdk
(\frac{1}{\Omega+i0+\varepsilon^{m}-\varepsilon^{m'}}-\frac{1}{\Omega+i0-\varepsilon^{m}+\varepsilon^{m'}})(1+{\rm cos}b)\\
=&\Sigma_{b1}^{m,intra}(\Omega,q)+\Sigma_{b2}^{m,intra}(\Omega,q),\\
\Sigma_{b}^{m,inter}(\Omega,q)&=
\frac{g_{s}g_{v}}{2}\int^{2\pi}_{0}d\theta\int^{\Lambda}_{k_{F}}kdk
(\frac{1}{\Omega+i0+\varepsilon^{m}+\varepsilon^{m'}}-\frac{1}{\Omega+i0-\varepsilon^{m}-\varepsilon^{m'}})(1-{\rm cos}b)\\
=&\Sigma_{b1}^{m,inter}(\Omega,q)+\Sigma_{b2}^{m,inter}(\Omega,q),
\end{aligned}
\end{equation}
where the first term of the intraband part can be obtained after some algebra as
\begin{equation} 
\begin{aligned}
\Sigma_{b1}^{m,intra}(\Omega,q)=&
k^{1-m}\sqrt{c_{1}^{2}+2c_{1}c_{2}k^{m}+(c_{2}^{2}+\xi^{2})k^{2m}}\\
&[
4k^{2}(m-1)F_{1}(\frac{3}{m}-1;-\frac{1}{2},-\frac{1}{2};\frac{3}{m};-\frac{k^{m}(c_{2}^{2}+\xi^{2})}{c_{1}c_{2}+\sqrt{-c_{1}^{2}\xi^{2}}}
,\frac{k^{m}(c_{2}^{2}+\xi^{2})}{-c_{1}c_{2}+\sqrt{-c_{1}^{2}\xi^{2}}})\\
&-(m-3)q^{2}{\rm sin}^{2}a F_{1}(\frac{1}{m}-1;-\frac{1}{2},-\frac{1}{2};\frac{1}{m}
;-\frac{k^{m}(c_{2}^{2}+\xi^{2})}{c_{1}c_{2}+\sqrt{-c_{1}^{2}\xi^{2}}},\frac{k^{m}(c_{2}^{2}+\xi^{2})}{-c_{1}c_{2}+\sqrt{-c_{1}^{2}\xi^{2}}})
]\\
&\frac{1}{2c_{1}c_{2}(m-3)(m-1)mq
\sqrt{\frac{-\sqrt{-c_{1}^{2}\xi^{2}}+c_{1}c_{2}+c_{2}k^{m}+\xi^{2}k^{m}}{c_{1}c_{2}-\sqrt{-c_{1}^{2}\xi^{2}}}}
\sqrt{\frac{ \sqrt{-c_{1}^{2}\xi^{2}}+c_{1}c_{2}+c_{2}k^{m}+\xi^{2}k^{m}}{c_{1}c_{2}+\sqrt{-c_{1}^{2}\xi^{2}}}}}\bigg|_{0}^{k_{F}},
\end{aligned}
\end{equation}
where $F_{1}(a;b,b';c;d,d')=\sum^{\infty}_{m,n=0}\frac{(a)_{m+n}(b)_{m}(b')_{n}}{m!n!(c)_{m+n}}d^{m}d'^{n}$ 
is the Appell hypergeometric function.
The other three terms can be obtained through the same way,
and then the
exchange-induced self-energy can also be obtained.
Note that
for the non-chiral Fermions, like the ones in 2D electron gas,
$\Sigma_{b}^{m,inter}(\Omega,q)$ vanishes since ${\rm cos}b=1$.

\subsection{3D Dirac system}

While for the 3D Dirac semimetals\cite{Yang B J,Xiong J,Wu Y} like the Na$_{3}$Bi or Cd$_{3}$As$_{2}$, PtTe$_{2}$,
the the chiral anomaly emerges since in odd space dimensions the anti-commute relation about the $\gamma_{5}$ matrix
is allowed, and
each Dirac node resolved into two Weyl nodes\cite{Xiong J} arrange along the $z$-direction of the momentum space
and with opposite chirality.
The Hamiltonian of the simplest 3D Dirac and Weyl semimetal are
$H=\sum_{i}\hbar v_{i}k_{i}\sigma_{i}$ ($i=x,y,z$)
and 
$H=\chi \sum_{i}\hbar v_{i}k_{i}\sigma_{i}+\chi \hbar v_{z}(k_{z}-\chi\delta k_{z})$ ($i=x,y$),
respectively.
The chiral effect gives the signs $\pm$ to $\xi$.
In 3D Dirac/Weyl semimetal,
the perturbations can remove the nodal line and leave the nodes\cite{Moors K},
while the nodes can not be removed but can only be shifted\cite{Burkov A A}.
As we mentioned above,
since the spin rotation is missing due to the Dirac $\delta$-type impurity field,
the rotational invariance is presented,
which is also partly due to the disorder averaging\cite{Bernardet K,John S},
and thus the disorder-induced self-energy is independent of the external momentum,
which reads
\begin{equation} 
\begin{aligned}
\Sigma^{D}(\omega)=\frac{n_{i}V_{0}^{2}}{\hbar^{2}}\int\frac{d^{3}k}{(2\pi)^{3}}\Gamma_{0}G_{0}^{mT}(\omega,k)\Gamma_{0},
\end{aligned}
\end{equation}
where $\Gamma_{0}$ is the irreducible vertex function which doesn't contains the Levi-Civita symbol here unlike the one in Ref.\cite{Isobe H}.
The vertex correction vanishes when it contains only the exchange-induced self-energy correction in instantaneous approximation,
which can be obtained by the Ward identity
$\frac{\partial \Sigma(\Omega,k)}{\partial \omega}=\Gamma(\Omega,k)$,
besides,
the vertex correction also vanishes in the large species-case (large $g$)
or when the integration momentum shell vanishes (the RG flow parameter $\ell=1$).
$n_{i}$ here is the impurity concentration,
$V_{0}$ is the impurity scattering potential (a scalar potential when only with the nonmagnetic impurity and without the magnetic impurity).

For the 3D Dirac/Weyl system,
we can write the Hamiltonian as
\begin{equation} 
\begin{aligned}
H_{0}^{mT}=\frac{k_{0}^{2-m}a}{2}\hbar v_{F} |{\bf k}|^{m}(\hat{k}\cdot\sigma)+\hbar v_{z}k_{z}\sigma_{3}
+(c_{1}+c_{2}|{\bf k}|^{m}\sigma_{z}\tau_{z})-\mu,
\end{aligned}
\end{equation}
the energy can be obtained as
\begin{equation} 
\begin{aligned}
\varepsilon^{mT}
=&-\mu+c_{1}+c_{2}|{\bf k}|^{m}\mp\\
&\sqrt{(\mu-(c_{1}+c_{2}|{\bf k}|^{m}))^{2}+\hbar^{2}v_{z}^{2}k_{z}^{2}+|{\bf k}|^{2m}\xi^{2}+2|{\bf k}|^{m}\xi{\rm cos}\theta(-\mu+(c_{1}+c_{2}|{\bf k}|^{m}))},\\
\approx & 
-\mu+c_{1}+c_{2}|{\bf k}|^{m}\mp
\sqrt{\hbar^{2}v_{z}^{2}k_{z}^{2}+|{\bf k}|^{2m}\xi^{2}},
\end{aligned}
\end{equation}
and the eigenvectors can still be written in a form similar to Eq.(38),
but the factor
$\alpha$ should be replaced by $\alpha^{T}={\rm arctan}\frac{\xi |{\bf k}|^{m}}{D^{T}}$
where $D^{T}$ is the gap opened by the collective effect of in-plane mass term ($c_{1}+c_{2}|{\bf k}|^{m}$) and the $z$-direction momentum.
Note that for the case of $\mu\neq 0$, the gap produced by the chemical potential should also be taken into account,
which should as large as $2|\mu|$.
And for the ultrathin film of Weyl semimetal, since the two Weyl nodes with opposite chirality are very close,
and
the $k_{z}$ is quantized\cite{Liu W E,Wang J},
which gives rise to a mass.

The velocity operators can be obtained by using the relation
$v_{\alpha}=\frac{\partial H_{0}^{m}}{\hbar \partial k_{\alpha}}$ ($\alpha=x,y,z$):
\begin{equation} 
\begin{aligned}
   %
v_{x}=
&\frac{1}{\hbar}\left[\begin{pmatrix}
m\xi |{\bf k}|^{m-1}e^{-i(m-1)\theta} & 0\\
0 & m\xi |{\bf k}|^{m-1} e^{i(m-1)\theta}
\end{pmatrix}
\sigma_{x}+mc_{2}|{\bf k}|^{m-1}{\rm cos}\theta\sigma_{3}\right]\\
=
&\frac{1}{\hbar}\begin{pmatrix}
mc_{2}|{\bf k}|^{m-1}{\rm cos}\theta & m\xi |{\bf k}|^{m-1}e^{-i(m-1)\theta} \\
 m\xi |{\bf k}|^{m-1} e^{i(m-1)\theta} & -mc_{2}|{\bf k}|^{m-1}{\rm cos}\theta
\end{pmatrix},\\
v_{y}=
&\frac{1}{\hbar}\left[\begin{pmatrix}
m\xi |{\bf k}|^{m-1} e^{-i(m-1)\theta} & 0\\
0 & m\xi |{\bf k}|^{m-1} e^{i(m-1)\theta}
\end{pmatrix}
\sigma_{y}+mc_{2}|{\bf k}|^{m-1}{\rm sin}\theta\sigma_{3}\right]\\
=
&\frac{1}{\hbar}\begin{pmatrix}
mc_{2}|{\bf k}|^{m-1}{\rm sin}\theta & -im\xi |{\bf k}|^{m-1}e^{-i(m-1)\theta} \\
 im\xi |{\bf k}|^{m-1} e^{i(m-1)\theta} & -mc_{2}|{\bf k}|^{m-1}{\rm sin}\theta
\end{pmatrix},\\
\end{aligned}
\end{equation}
where the relation $\frac{\partial k}{\partial k_{\alpha}}=\frac{k_{\alpha}}{k}$ ($\alpha=x,y,z$) is used.
Here we write the third Pauli matrix as $\sigma_{3}$ to distinguish from the spin operator in $z$ direction.
The velocity matrix elements can be obtained base on the above velocity operators,
\begin{equation} 
\begin{aligned}
  %
\langle \varepsilon |v_{x}|\varepsilon'\rangle=&
\mp \frac{|{\bf k}|^{m-1}m\sqrt{c_{2}^{2}+2\xi^{2}+c_{2}^{2}{\rm cos}(2\theta)}}
{\sqrt{2}}(1-\delta_{s,s'}),\\
\langle \varepsilon |v_{y}|\varepsilon'\rangle=&
\mp \frac{i|{\bf k}|^{m-1}m\sqrt{-c_{2}^{2}-2\xi^{2}+c_{2}^{2}{\rm cos}(2\theta)}}
{\sqrt{2}}(1-\delta_{s,s'}).\\
\end{aligned}
\end{equation}

Considering only the spin-flipping during the scattering,
the optical conductivity can be obtained as $\sigma(\omega)=\sigma_{ij}+\sigma_{ij}'$,
with
\begin{equation} 
\begin{aligned}
\sigma_{ij}=&-i\hbar e^{2}
  \int\frac{d^{3}k}{(2\pi)^{3}}
\frac{N_{F}(\varepsilon_{-k})-N_{F}(\varepsilon_{+k})}{\varepsilon_{-k}-\varepsilon_{+k}}
\frac{\langle \varepsilon'|v_{i}|\varepsilon\rangle\langle\varepsilon|v_{j}|\varepsilon'\rangle}
{\hbar \omega-\varepsilon_{+k}+\varepsilon_{-k}+i\eta},
\end{aligned}
\end{equation}
for scattering from up-spin to down-spin,
and 
\begin{equation} 
\begin{aligned}
\sigma_{ij}'=&-i\hbar e^{2}
  \int\frac{d^{3}k}{(2\pi)^{3}}
\frac{N_{F}(\varepsilon_{+k})-N_{F}(\varepsilon_{-k})}{\varepsilon_{+k}-\varepsilon_{-k}}
\frac{\langle \varepsilon|v_{i}|\varepsilon'\rangle\langle\varepsilon'|v_{j}|\varepsilon\rangle}
{\hbar \omega-\varepsilon_{-k}+\varepsilon_{+k}+i\eta},
\end{aligned}
\end{equation}
for scattering from down-spin to up-spin.
Then we can obtain
\begin{equation} 
\begin{aligned}
                                   \sigma_{xy}
=&
\frac{-i\hbar e^{2}}{(2\pi)^{3}}
  \frac{1.5708i km^{2}}
       {c_{2}^{2}(i\eta+\hbar\omega)}
  \sqrt{c_{2}^{2}+2\xi^{2}+c_{2}^{2}{\rm cos}(2\theta)}
  \sqrt{-c_{2}^{2}-2\xi^{2}+c_{2}^{2}{\rm cos}(2\theta)}\\
&\left[
 -i\eta -\hbar\omega+2c_{1}{}_{2}F_{1}[1,\frac{1}{m};1+\frac{1}{m};-\frac{c_{2}k^{m}}{c_{1}}]\right.\\
&\left.+(i\eta-2c_{1}+\hbar\omega){}_{2}F_{1}[1,\frac{1}{m};1+\frac{1}{m};-\frac{c_{2}k^{m}}{-5i\eta+c_{1}-0.5\hbar\omega}]
\right]\bigg|_{k},
\end{aligned}
\end{equation}
\begin{equation} 
\begin{aligned}
                                       \sigma_{xx}
=&
\frac{-i\hbar e^{2}}{(2\pi)^{3}}
  \frac{1.5708 km^{2}}
       {c_{2}^{2}(i\eta+\hbar\omega)}
(c_{2}^{2}+2\xi^{2}+c_{2}^{2}{\rm cos}(2\theta))\\
&\left[
 -i\eta -\hbar\omega+2c_{1}{}_{2}F_{1}[1,\frac{1}{m};1+\frac{1}{m};-\frac{c_{2}k^{m}}{c_{1}}]\right.\\
&\left.+(i\eta-2c_{1}+\hbar\omega){}_{2}F_{1}[1,\frac{1}{m};1+\frac{1}{m};-\frac{c_{2}k^{m}}{-5i\eta+c_{1}-0.5\hbar\omega}]
\right]\bigg|_{k},
\end{aligned}
\end{equation}
\begin{equation} 
\begin{aligned}
                                            \sigma_{yy}
=&
\frac{-i\hbar e^{2}}{(2\pi)^{3}}
  \frac{-1.5708 km^{2}}
       {c_{2}^{2}(i\eta+\hbar\omega)}
(-c_{2}^{2}-2\xi^{2}+c_{2}^{2}{\rm cos}(2\theta))\\
&\left[
 -i\eta -\hbar\omega+2c_{1}{}_{2}F_{1}[1,\frac{1}{m};1+\frac{1}{m};-\frac{c_{2}k^{m}}{c_{1}}]\right.\\
&\left.+(i\eta-2c_{1}+\hbar\omega){}_{2}F_{1}[1,\frac{1}{m};1+\frac{1}{m};-\frac{c_{2}k^{m}}{-5i\eta+c_{1}-0.5\hbar\omega}]
\right]\bigg|_{k}.
\end{aligned}
\end{equation}
Similarly, the $\sigma_{ij}'$ can be obtained in the same way.
The results of the optical conductivity are presented in the Fig.1-3,
where we can see that for optical transition from up-spin to down-spin (Fig.1), or the total one (Fig.3),
the longitudinal and transverse conductivities are linear with frequency $\omega$ in the low-energy region,
which is consistent with the result of Ref.\cite{Ahn S}.
In Fig.1, the slope of such linear conductivity can be approximately written as $10m\frac{i\hbar e^{2}}{(2\pi)^{3}}$,
while that of the total optical conductivity is independent of $m$. 
From Fig.1, we can also see that the optical conductivity increases with photon frequency and then saturates in large value of $\omega$.
For larger $m$, the saturation happen earlier.
The effect of angle $\theta$ (i.e., the degree of anisotropy)
is enhanced with the increase of $m$.
But from Fig.3, we can see that for transverse total optical conductivity,
the effect of anisotropy is supressed.
The total optical conductivity is proportional to the order $m$.
We also find that, for isotropic case ($\theta=\Phi=\pi/4$),
$\sigma_{xx}=\sigma_{yy}$, which implies that the system has 4-fold rotational symmetry ($C_{4}$ symmetry) in isotropic case
no matter how large the $m$ is.

\subsection{3D Weyl system}

The Dirac nodes can be divided into the Weyl nodes along the $z$-direction by using the projection operator\cite{Isobe H},
and the Hamiltonian with topological winding number $\chi$ (chirality) reads
\begin{equation} 
\begin{aligned}
H_{0}^{mT}=\frac{k_{0}^{2-m}a}{2}\hbar v_{F} k^{m}(\hat{k}\cdot\sigma)
+\chi\hbar v_{z}(k_{z}-\chi \delta k_{z})\sigma_{z}+(c_{3}+c_{4}(k_{z}-\chi \delta k_{z})^{n})\sigma_{z}\tau_{z}-\mu_{\chi},
\end{aligned}
\end{equation}
where $\mu_{\chi}$ is the chemical potential with chirality $\chi=\pm 1$
and $\mu_{+}=\mu_{-}=\mu$ in undoped case.
$\delta k_{z}$ is the distance in momentum space removed from the previous Dirac node
which explicitly breaks the time-reversal symmetry.
$v_{z}=\frac{at_{\perp}{\rm sin}(\delta k_{z}a)}{\hbar}$ is the $z$-direction velocity.
Here we define $k_{x}=k{\rm cos}\theta$, $k_{y}=k{\rm sin}\theta$, $k={\bf k}{\rm sin}\varphi$,
$k_{z}={\bf k}{\rm cos}\varphi$,
and still use the defination $\frac{k_{0}^{2-m}a}{2}\hbar v_{F}=\xi$.
The first term of the above Hamiltonian contains no out-of-plane components,
which indicates the untilted type-I Weyl semimetal
when the mass term is missing.
The mass term is dominated by the momentum $k_{z}$ here rather than the in-plane momentum as shown in the previous model
and it explicitly breaks the inversion symmetry.
Then eigenvalues can be obtained as
\begin{equation} 
\begin{aligned}
\varepsilon^{mT}=&-\mu-\hbar v_{z}(\delta k_{z}-\chi k_{z})\\
&\mp
\sqrt{[c_{3}+c_{4}(k_{z}-\chi\delta k_{z})^{n}]^{2}+k^{2m}\xi^{2}
+(\delta k_{z}-k_{z}\chi)^{2}-2\eta k^{m}\xi(\mu+\hbar v_{z}(\delta k_{z}-k_{z}\chi)){\rm cos}\theta}\\
\approx &-\mu-\hbar v_{z}(\delta k_{z}-\chi k_{z})\mp
\sqrt{[c_{3}+c_{4}(k_{z}-\chi\delta k_{z})^{n}]^{2}+k^{2m}\xi^{2}}.
\end{aligned}
\end{equation}
The imaginary part and real part of the bare Green's function (Fermion propagator) $G_{0}^{mT}(\omega,k)$ are
\begin{equation} 
\begin{aligned}
{\rm Im}G_{0}^{mT}(\omega,k)=&-\pi\delta(\omega-\varepsilon^{mT}),\\
{\rm Re}G_{0}^{mT}(i\omega,k)=&\frac{2}{\pi}\int^{\infty}_{0}d\omega\frac{\omega}{\omega^{2}-(i\omega)^{2}}{\rm Im}G_{0}^{mT}(\omega,k)\\
=&\frac{-2\varepsilon^{mT}\theta(\varepsilon^{mT})}{(\varepsilon^{mT})^{2}+\omega^{2}},
\end{aligned}
\end{equation}
where the Sokhotski-Plemelj theorem and Kramers-Kronig relation are used.
$\theta(x)$ here is the step function.

Using the eigenvalue
\begin{equation} 
\begin{aligned}
\varepsilon^{mT}=-\mu-\hbar v_{z}(\delta k_{z}-\chi (k{\rm cos}\varphi))\mp
\sqrt{c_{3}^{2}+(k{\rm sin}\varphi)^{2m}\xi^{2}},
\end{aligned}
\end{equation}
where we further set $c_{4}=0$,
then
after some algebra, the above disorder-induced self-energy can be obtained as
\begin{equation} 
\begin{aligned}
\Sigma^{D}(\omega)=&\frac{n_{i}V_{0}^{2}}{\hbar^{2}}
\int^{2\pi}_{0}d\theta\int^{\pi}_{0}\frac{1}{m}({\rm sin}\varphi)^{2-m} d\varphi\int^{\Lambda}_{0}dk
\ k^{2}\Gamma_{0}G_{0}^{mT}(\omega,k)\Gamma_{0}\\
=&\frac{2\pi n_{i}V_{0}^{2}}{m\hbar^{2}}
\int^{\pi}_{0}({\rm sin}\varphi)^{2-m} d\varphi\\
&\frac{1}{12}\Lambda^{3}
\left\{\frac{-4\delta k_{z}\hbar v_{z}-
[
4c_{3}^{2}m\sqrt{\frac{\xi^{2}(\Lambda{\rm sin}\varphi)^{2m}}{c_{3}^{2}}+1}
{}_{2}F_{1}(\frac{1}{2},\frac{3}{2m};1+\frac{3}{2m};-\frac{\xi^{2}(\Lambda{\rm sin}\varphi)^{2m}}{c_{3}^{2}})
]}
{[(m+3)\sqrt{c_{3}^{2}+\xi^{2}(k{\rm sin}\varphi)^{2m}}]}\right.\\
&\left.-\frac{12\sqrt{c_{3}^{2}+\xi^{2}(\Lambda{\rm sin}\varphi)^{2m}}}{m+3}
+3\chi\hbar v_{z}\Lambda{\rm cos}\varphi-4\mu
\right\},
\end{aligned}
\end{equation}
where ${}_{2}F_{1}(a,b;c;d)=\sum^{\infty}_{0}\frac{(a)_{n}(b)_{n}}{(c)_{n}}\frac{d^{n}}{n!}$ is the hypergeometric function.
In such a 3D Dirac system,
the Fermion spectral function in the absence of the quasiparticle scattering (noninteracting case) reads
\begin{equation} 
\begin{aligned}
A^{mT}_{0}=\delta(\omega-\varepsilon^{mT}),
\end{aligned}
\end{equation}
where the locations of the sharp peaks are surely related to the above band dispersion.
The longitudinal optical conductivity $\sigma_{zz}$ is finite unless when a pair of Weyl nodes tilted in parallel direction\cite{Das K}
in momentum space.
The density of states for the 3D Dirac semimetal at half-filling can then be obtained as
\begin{equation} 
\begin{aligned}
D^{mT}=&\int\frac{d^{3}k}{(2\pi)^{3}}A_{0}^{mT}\\
=&\int\frac{d^{3}k}{(2\pi)^{3}}\delta(\omega-(\hbar v_{z}k_{z}\mp k^{m}\xi))\\
\approx &\frac{1}{\pi}\int dk_{z}\int dk \frac{c}{c^{2}+(\hbar v_{z}k_{z}\mp k^{m}\xi)^{2}}\\
=&{\rm const.}_{1}+{\rm const.}_{2}k\\
&-\frac{ikm}{2\hbar v_{z}}
[
(\frac{\xi km}{-\hbar v_{z}k_{z}+\xi k^{m}-ic})^{-1/m}
{}_{2}F_{1}(-\frac{1}{m},-\frac{1}{m};\frac{m-1}{m};\frac{c-i\hbar v_{z}k_{z}}{i\xi k^{m}+c-i\hbar v_{z}k_{z}})\\
&-
(\frac{\xi km}{-\hbar v_{z}k_{z}+\xi k^{m}+ic})^{-1/m}
{}_{2}F_{1}(-\frac{1}{m},-\frac{1}{m};\frac{m-1}{m};\frac{c+i\hbar v_{z}k_{z}}{-i\xi k^{m}+c+i\hbar v_{z}k_{z}})
]\\
&-\frac{k{\rm tan}^{-1}(\frac{\xi k^{m}-\hbar v_{z}k_{z}}{c})}{\hbar v_{z}},
\end{aligned}
\end{equation}
where $c$ is the small quantity used in the Lorentzian representation.

Base on the spectral function in the absence of the self-energy correction,
the optical conductivity per Weyl node in the presence of nonzero Fermionic Matsubara frequency can be obtained by the Kubo formula
\begin{equation} 
\begin{aligned}
\sigma_{zz}(\omega)=\frac{{\rm Im}\Pi_{zz}(\omega+i0)}{\omega},
\end{aligned}
\end{equation}
then we have
\begin{equation} 
\begin{aligned}
{\rm Re}\sigma_{zz}(\omega)=&e^{2}\int^{\infty}_{-\infty}\frac{d\Omega}{2\pi}\int\frac{d^{3}k}{(2\pi)^{3}}
\frac{N_{F}(\Omega-\mu)-N_{F}(\Omega'-\mu)}{\Omega'-\Omega+\omega+i0}{\rm Tr}[\hat{v_{z}}A^{mT}_{0}(\Omega',k)\hat{v_{z}}A^{mT}_{0}(\Omega,k)]\\
=&\frac{e^{2}}{2\omega+2i0}\int^{\mu}_{\mu-\omega}\frac{d\Omega}{2\pi}\int\frac{d^{3}k}{(2\pi)^{3}}
v_{z}^{2}
\begin{pmatrix}
A^{mT}_{1}(\Omega')-A^{mT}_{3}(\Omega') & A^{mT'}_{2}(\Omega')-A^{mT'}_{4}(\Omega')\\
A^{mT}_{2}(\Omega)-A^{mT}_{4}(\Omega) & A^{mT'}_{1}(\Omega)-A^{mT'}_{3}(\Omega)\\
\end{pmatrix},
\end{aligned}
\end{equation}
where we define $\Omega'=\Omega+\omega$,
$A^{mT'}$ is the spectral function after optical transition.
The velocity operator is $\hat{v_{z}}=v_{z}\sigma_{z}$.
The components of the spectral function read
\begin{equation} 
\begin{aligned}
A^{mT}_{1}(\Omega)=&\pi[\delta(\Omega+\varepsilon^{mT})+\delta(\Omega-\varepsilon^{mT})],\\
A^{mT}_{3}(\Omega)=&\pi[\delta(\Omega+\varepsilon^{mT})-\delta(\Omega-\varepsilon^{mT})],\\
A^{mT}_{2}(\Omega)=&\pi[-\delta(\Omega+\varepsilon^{mT})-\delta(\Omega-\varepsilon^{mT})]{\rm sgn}(\varepsilon^{mT}),\\
A^{mT}_{4}(\Omega)=&\pi[-\delta(\Omega+\varepsilon^{mT})+\delta(\Omega-\varepsilon^{mT})]{\rm sgn}(\varepsilon^{mT}),\\
\end{aligned}
\end{equation}
then the above optical conductivity can be rewritten as
\begin{equation} 
\begin{aligned}
{\rm Re}\sigma_{zz}(\omega)
=&\frac{e^{2}}{2\omega+2i0}\int^{\mu}_{\mu-\omega}\frac{d\Omega}{2\pi}\int\frac{d^{3}k}{(2\pi)^{3}}
4v_{z}^{2}
\begin{pmatrix}
\pi\delta(\Omega'-\varepsilon^{mT}) & -\pi\delta(\Omega'-\varepsilon^{mT'}){\rm sgn}(\varepsilon^{mT'})\\
-\pi\delta(\Omega-\varepsilon^{mT}){\rm sgn}(\varepsilon^{mT}) & \pi\delta(\Omega-\varepsilon^{mT})
\end{pmatrix}\\
=&\frac{e^{2}}{2\omega+2i0}\int^{\mu}_{\mu-\omega}\frac{d\Omega}{2\pi}\int\frac{d^{3}k}{(2\pi)^{3}}
4v_{z}^{2}
\pi^{2}
\delta(\Omega-\varepsilon^{mT})
(\delta(\Omega'-\varepsilon^{mT})-\delta(\Omega'-\varepsilon^{mT'}){\rm sgn}(\varepsilon^{mT}){\rm sgn}(\varepsilon^{mT'}))\\
=&\frac{2v^{2}_{z}\pi^{2}e^{2}}{\omega+i0}\int^{\mu}_{\mu-\omega}\frac{d\Omega}{2\pi}
\frac{1}{m}\int^{2\pi}_{0}d\theta\int^{\pi}_{0}d\varphi({\rm sin}\varphi)^{2-m}\int^{\Lambda}_{0}dk\\
&k^{2}
\delta(\Omega-\varepsilon^{mT})
(\delta(\Omega'-\varepsilon^{mT})-\delta(\Omega'-\varepsilon^{mT'}){\rm sgn}(\varepsilon^{mT}){\rm sgn}(\varepsilon^{mT'})).\\
\end{aligned}
\end{equation}
$\varepsilon^{mT'}$ is the eigenenergy after the optical transition,
and we note that, since we assume the hopping strength is isotropic and real (i.e., does not consider the chiral kind of hopping),
the eigenenergy (dispersion) is real.
Then the above integral can be solved by using the Lorentzian representation $\delta(x)=\frac{c}{\pi(c^{2}+x^{2})}$ again
where $c$ is a small quantity related to the quaisparticle scattering.

The particle spectral function including the disorder-induced self-energy effect reads\cite{Wang J R,22,Hassaneen K S A}
\begin{equation} 
\begin{aligned}
A^{mT}(\Omega)=-\frac{1}{\pi}\frac{|{\rm Im}\Sigma^{D}(\Omega)|}{(\Omega-{\rm Re}\Sigma^{D}(\Omega)-\varepsilon^{mT})^{2}+({\rm Im}\Sigma^{D}(\Omega))^{2}},
\end{aligned}
\end{equation}
which contains the informations about not only the dispersion but also the
quasiparticle residue and the Fermion relaxation.
In the presence of the screened long-range Coulomb interaction
by the collective excitations in non-Fermi-liquid state (but with finite chemical potential),
the above perturbed spectral function also related to the excitation damping,
like the plasmon mode which damped into the particle-hole excitations due to the non-zero 
imaginary part of the polarization function (Bosonic self-energy) as we studied\cite{1,2,3,4}.
The hole spectral function can be obtained simply as $-A^{mT}(\Omega)$.

\section{Conclusion}

In conclusion,
we investigate the self-energy correction,
symmetry,
free energy,
transverse optical conductivity
of the 2D Dirac system in non-Fermi-liquid state.
The non-Fermi-liquid behaviors of the 2D and 3D Dirac/Weyl systems with higher order dispersion
are also studied and we found that the non-Fermi-liquid features remain even at finite chemical potential,
and they are distinct from the Fermi-liquid picture and the conventional non-Fermi-liquid picture.
In the presence of the impurity scattering, the Fermionic/Bosonic polaron formed by dressing the Fermion/Boson majority particles
as widely found in ultracold Fermi gases\cite{Schirotzek A,Kohstall C} and BEC\cite{Li W}, respectively,
and they are also important in studying the 
perturbation effect within the contact potential (Dirac $\delta$-type
impurity field) context.
The many-body perturbation effect at the charge neutrality (the critical point) 
is related to the long-range Coulomb interaction and the coupling to the electronic critical mode\cite{Han S E}.
Thus it is also interesting to considering the results in this paper to the polaron dynamics\cite{wums19,wums20}.
In the presence of the gapless order parameter fluctuation,
the Landau damping of the longitudinal excitations within RPA for the non-Fermi-liquid case
are also discussed.
For Weyl semimetal,
the Chern number is given by $\chi m$, and
the momentum $k_{z}$ along the diraction of the line connecting two Weyls nodes
can be related to the topological phase transition,
like the transition between trivial insulator phase and the quantum anomalous Hall phase.
Further, since the higher order dispersion (with $m>1$),
the DOS remains finite at zero-energy limit (charge neutrality),
that makes the phase transition induced by interaction at this point more easy\cite{Nandkishore R}.
Further, for heavily tilted Weyl semimetal (i.e., the type-II where the tilt velocity is larger than the 
planar one and with unclosed electron pocket)\cite{Farajollahpour T,DD},
the occupied states along the $k_{z}$ may create the particle-hole pairs\cite{Zubkov M A} (particle-hole fluctuation),
and physical observables in non-Fermi liquid picture like optical conductivity\cite{Ahn S}, can also be affected by the tilt.

\section{Appendix.A}

The Feynman diagrams of fermionic and bosonic self-energy are 
shown in Fig.4, where the external propagators are attached with
the fermion loop.
We note that, in the presence of electromagnetic field (like a gate voltage),
the electromagnetic response can also be described by the Fermion loop as shown in Fig.4(b),
with the electron propagator $G_{0}$ and external photon propagator\cite{Isobe H,Jalali-Mola Z,Pozo ?}.

 For 2D Dirac system, the Coulomb induced fermi exchange self-energy for high-order dispersion reads
\begin{equation} 
\begin{aligned}
\Sigma_{C}(k,\omega)&=\int^{\infty}_{-\infty}\frac{d\Omega}{2\pi}\int\frac{d^{2}q}{(2\pi)^{2}}G_{0}(k+q,\omega+\Omega)\frac{2\pi e^{2}}{\epsilon_{0}}\frac{1}{q-\frac{2\pi e^{2}}{\epsilon_{0}}\Pi(q,\Omega)},\\
\Pi(q,\Omega)&=g \int^{\infty}_{-\infty}\frac{d\omega}{2\pi}\int\frac{d^{2}k}{(2\pi)^{2}}
{\rm Tr}[G_{0}(k+q,\omega+\Omega)G_{0}(k,\omega)],
\end{aligned}
\end{equation}
where the $1/q$ in the denominator of the self-energy is the bare boson propagator $D_{0}$
which corresponds to the GV approximation, when the dielectric term
(i.e., the dynamical polarization or density-density response function) is contained,
the dressed boson propagator $D=\frac{1}{q(1-\frac{2\pi e^{2}}{\epsilon_{0}q}\Pi(q,\Omega))}$ corresponds to the GW approximation,
as despicted by the wave line in Fig.9(a).
$\epsilon_{0}$ is the background dielectric constant.
Here the bare propagator is approximated as $D_{0}=1/q$,
which is similar to the one $D'_{0}=1/q^{2}$ for three-dimensional case\cite{Goswami P},
however, the bare bosonic propagators may have more complex form in scaling analysis\cite{Metlitski M A,González J}.

Then we focus on the regime where $q,\Omega\rightarrow 0$.
For approximated energy as stated above,
we have $\varepsilon^{m}=\sqrt{\xi^{2}k^{2m}+D^{2}}-\mu$ 
where $D$ denotes the band gap which is momentum-independent (due to the long-wave length and low-energy limit).
%
Then the polarization function (density-density response or equivalently, the bosonic self-energy) 
 at dynamical limit to order of $q^{2}$ reads\cite{Sachdeva R,Sarma S D}
\begin{equation} 
\begin{aligned}
\Pi(q,\Omega\rightarrow 0)=\frac{gv_{F}}{4\pi\hbar\Gamma(2)}\frac{\xi^{2}k_{F}^{2}}{\sqrt{\xi^{2}k_{F}^{2}+D^{2}}}\frac{q^{2}}{\Omega^{2}},
\end{aligned}
\end{equation}
where $g$ is the degenerate number, $k_{F}=\sqrt{\mu^{2}-D^{2}}$,
and the carrier concentration reads
$n_{c}=\frac{g\pi k_{F}^{2}}{4\pi^{2}\Gamma(2)}$.
This expression implies that the polarization is independent of the order $m$ in the long-wave length and low-energy limit,
which will be verified in the following.
Although
this approximated polarization function inevitably introduce the pole at static limit $\Omega=0$,
it will be proved to be correct 
(in following) for arbitary order $m$ in the limit of $q\rightarrow 0$ and $\Omega\rightarrow 0$.
The long-wave length limit usually related to the Fermi liquid picture with $q\ll k_{F}$\cite{Raghu S},
which corresponds to the local density response.
As can be seen, the above exchange self-energy does not contains the ladder correction.
We can also easily see that there is a logarithmic divergence,
which is consistent with the result of Ref.\cite{Vafek O},
such logarithmic divergence will appears also in the disorder-induced self-energy,
but unlike the result in Hartree-Fock approximation\cite{Vafek O,Elias D C}:
$\Sigma_{C}\sim\frac{e^{2}}{4\epsilon_{0}}{\rm ln}\frac{\Lambda}{k}$
which is frequency-independent,
the exchange self-energy here will becomes zero in the static limit due to the inclusion of dielectric term.

Using the approximation up to order of $q^{2}$
\begin{equation} 
\begin{aligned}
\varepsilon_{k}-\varepsilon_{k'}=&-\frac{\xi^{2}q k^{m}}{\varepsilon_{k}}-\frac{q^{2}D^{2}\xi^{2}}{2(\varepsilon_{k}^{2})^{3/2}}+O(q^{3}),\\
\varepsilon_{k}\varepsilon_{k'}=&(\varepsilon_{k}^{2})+\xi^{2}q k^{m}  +\frac{D^{2}\xi^{2}q^{2}}{2\varepsilon_{k}^{2}} +O(q^{3}),
\end{aligned}
\end{equation}
the bosonic self-energy can be obtained by the Lehmann type formular as
\begin{equation} 
\begin{aligned}
\Pi(q,\Omega)=&
2\pi\sum_{ss'}
\int^{\Lambda}_{0}k g\frac{1}{2}(1+ss'{\rm cos}b+\frac{D^{2}}{s\varepsilon_{k}s'\varepsilon_{k'}})
\frac{1}{\Omega+i0+s\varepsilon_{k}-s'\varepsilon_{k'}}dk,
\end{aligned}
\end{equation}
which for intraband transition ($ss'=1$) equivalents to
\begin{equation} 
\begin{aligned}
\Pi(q,\Omega)=&
\frac{0.00316629 gk^{2-3m}(q\sqrt{\xi^{2}k^{2m}}+k^{m}(\Omega+i\eta))(4k^{2m}(-1+m)+q^{2}{\rm sin}^{2}a)}
{(-1+m)(\Omega^{2}-q^{2}\xi^{2}+2i\Omega\eta-\eta^{2})}\\
&+\frac{0.00316629 gk^{2-3m}(-q\sqrt{\xi^{2}k^{2m}}+k^{m}(\Omega+i\eta))(4k^{2m}(-1+m)+q^{2}{\rm sin}^{2}a)}
{(-1+m)(\Omega^{2}-q^{2}\xi^{2}+2i\Omega\eta-\eta^{2})}\bigg|_{0}^{\Lambda},
\end{aligned}
\end{equation}
where the first term corresponds to the intra-conduction band transition 
and the second term corresponds to the intra-valence band transition,
and we set $D=0$ and $\mu=0$ here for simplicity.
From the this expression, we can know that in static limit ($\Omega=0$) the total intraband polarization vanishes,
and the polarization function obtained here in this approximation can not be applied to the case of linear dispersion $m=1$.
The divergence of $\Pi(q,\Omega)$ can be seen from the first column of Fig.5,
where we put it in the long-wave length limit first and then lowers the frequency to zero,
which is in agree with Eq.(70).
But we note that, when $q=\Omega=0$,
$\Pi=0$.
While for the interband part of the polarization $(ss'=-1)$,
the result is much more complicated than the intraband part
(contains a series of hypergeometric functions), and an analytical solution is impossible to obtained until we
set $m\ge 3$.
That also reveals great difference compared to the ordinary Fermi liquid or the non-chiral electron gas.
From Fig.6, we can see that the imaginary part of the polarization in static limit is still nearly independent of the order $m$,
and the rest panels show that, the imaginary part is nonzero only around the point where $\Omega=v_{F}k$,
which is due to the vanishing particle-hole continuum region,
and we find the linewidth of the peak increases with the increase of $m$.

The self-energy and spectral function about the intra-conduction band transition
are presented in the Fig.7 and Fig.8, respectively.
Note that the self-energy here is calculated in long-wavelength limit $q\rightarrow 0$,
which can be recognized as $q\ll t $ in the lattice model where $t$ is the hopping amplitude.
The spectral function in Fig.8 is calculated according to the retarded self-energy (Fig.6-7)
based on the dressed boson propagator.
We can see that the width of imaginary part of self-energy decrease with the increase of order $m$,
and their locations also shift toward right with the increase of electron frequency,
but the shift distance decreases with the increase of $m$.
Similar phenomenon can be seen from the 
spectral functions which all have finite peak width,
and that is different to the Fermi liquid in normal metals where
the shape of spectral function is close to the $\delta$ function due to the weak interacting nature.
Except the static case, we find that the peaks in spectral function are symmetrical.
Then base on the spectral function,
the DOS as well as the occupation probability can be obtained by integrate over the momentum $k$ and $\omega$,
respectively.

\end{large}
\renewcommand\refname{References}

\clearpage

Fig.1
\begin{figure}[!ht]
   \centering
 \centering
   \begin{center}
     \includegraphics*[width=1\linewidth]{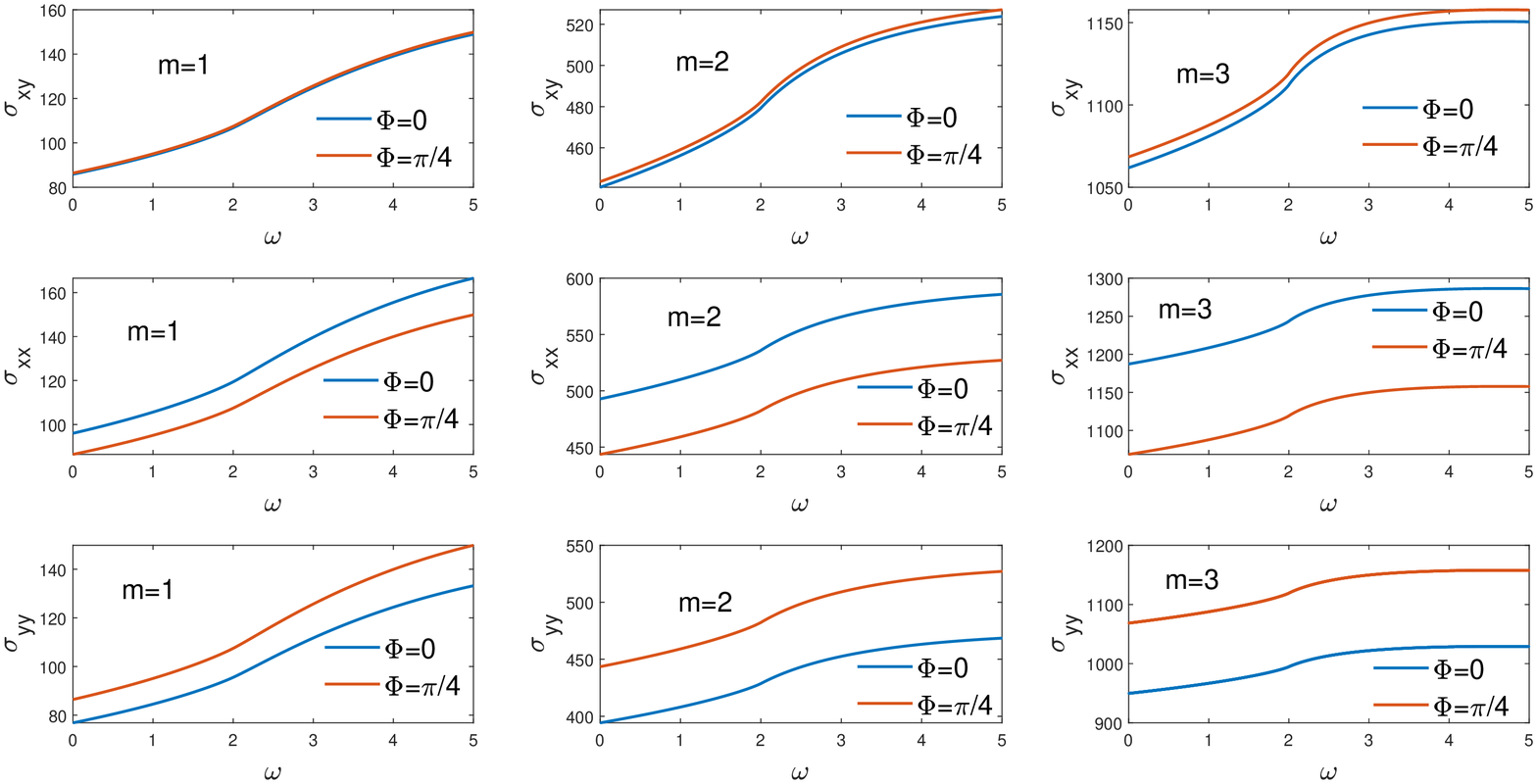}
\caption{Opticl conductivity in 3D Dirac system for interband scattering from up-spin to down-spin.
The first row is the transverse conductivity $\sigma_{xy}$.
The second and third rows are the longitudinal conductivities $\sigma_{xx}$ and $\sigma_{yy}$.
The colums from left to right correspond to the order $m=1,\ 2$ and 3.
The vertical axis is in unit of $\frac{i\hbar e^{2}}{(2\pi)^{3}}$.
Here we set $c_{1}=c_{2}=1$, $\xi=2$.
Since we consider the low-temperature limit and clean limit, and set chemical potential.as $\mu=0$,
 the optical conductivities here don't show a gap due to the Pauli blocking.
}
   \end{center}
\end{figure}
\clearpage

Fig.2
\begin{figure}[!ht]
   \centering
 \centering
   \begin{center}
     \includegraphics*[width=1\linewidth]{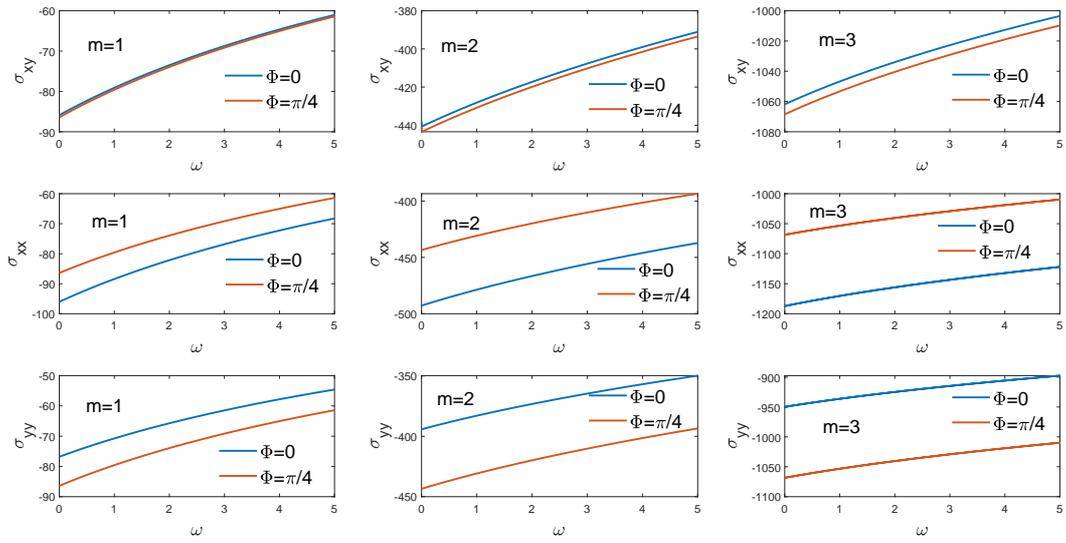}
\caption{The same as Fig.2 but for interband scattering from down-spin to up-spin.
}
   \end{center}
\end{figure}
\clearpage

Fig.3
\begin{figure}[!ht]
   \centering
 \centering
   \begin{center}
     \includegraphics*[width=1\linewidth]{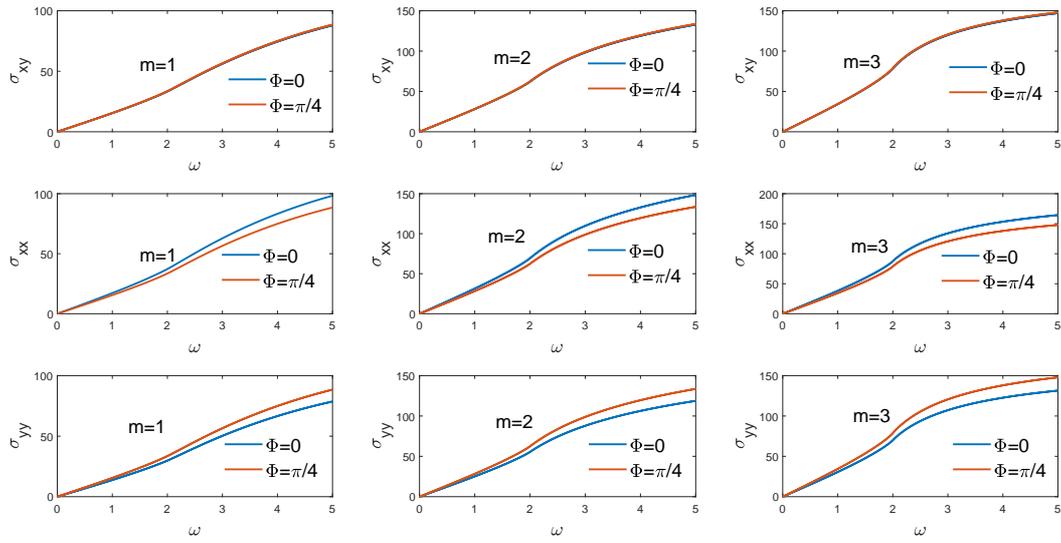}
\caption{The total optical conductivity where we consider the interband scatterings from down-spin to up-spin
and up-spin to down-spin.
}
   \end{center}
\end{figure}

\clearpage

Fig.4
\begin{figure}[!ht]
   \centering
 \centering
   \begin{center}
     \includegraphics*[width=1\linewidth]{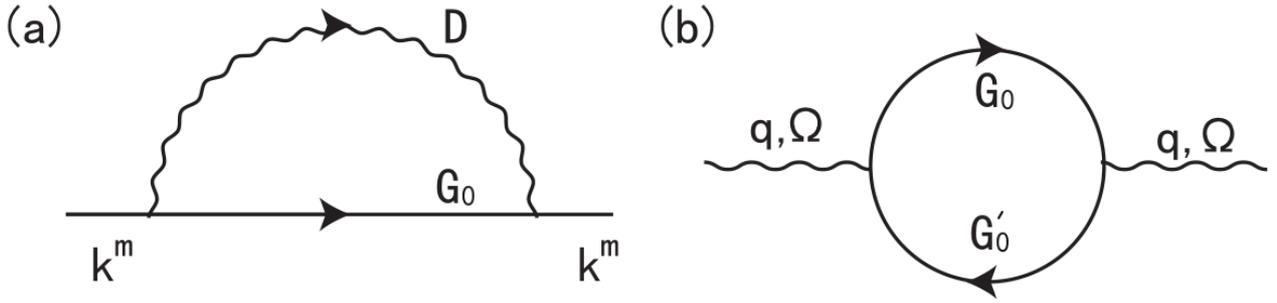}
\caption{
(a) Fermion self-energy with the dressed scalar potential propagator $D$.
(b) Boson self-emergy with scattering momentum and frequency $q,\Omega$.
}
   \end{center}
\end{figure}

Fig.5
\begin{figure}[!ht]
   \centering
 \centering
   \begin{center}
     \includegraphics*[width=1\linewidth]{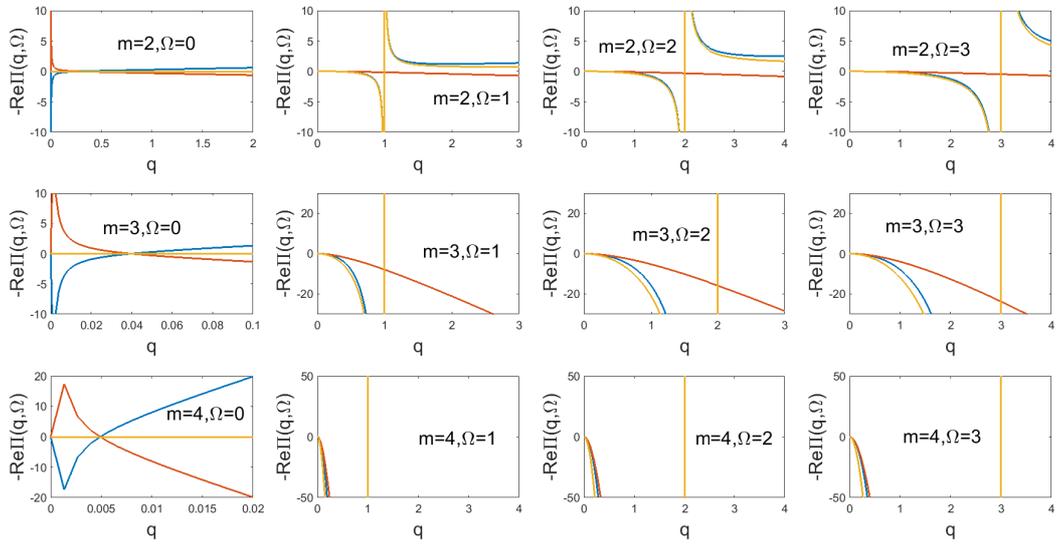}
\caption{Intra-conduction band part (blue) intra-valence band part (red) of the polarization function,
and their summation (yellow).
The vertical axis is in unit of $2\pi$
and we set $a=\pi/4$, $\xi=1$.
}
   \end{center}
\end{figure}
\clearpage

Fig.6
\begin{figure}[!ht]
   \centering
 \centering
   \begin{center}
     \includegraphics*[width=1\linewidth]{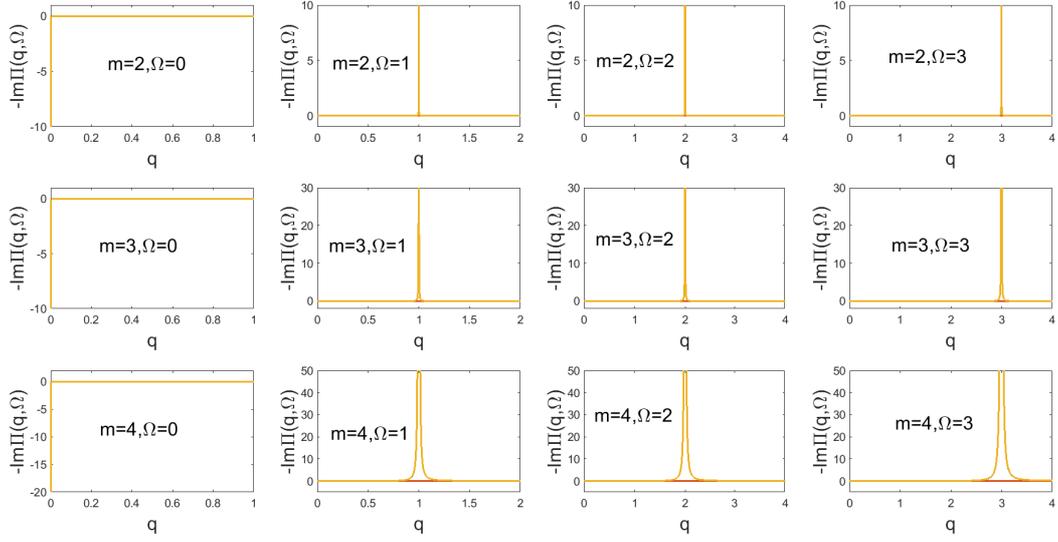}
\caption{Imaginary part of the polarization function corresponds to Fig.5.
We can see that the imaginary part of the polarization (and the spectral function) only contributed by the 
 intra-conduction band part (blue lines).
}
   \end{center}
\end{figure}
Fig.7
\begin{figure}[!ht]
   \centering
 \centering
   \begin{center}
     \includegraphics*[width=1\linewidth]{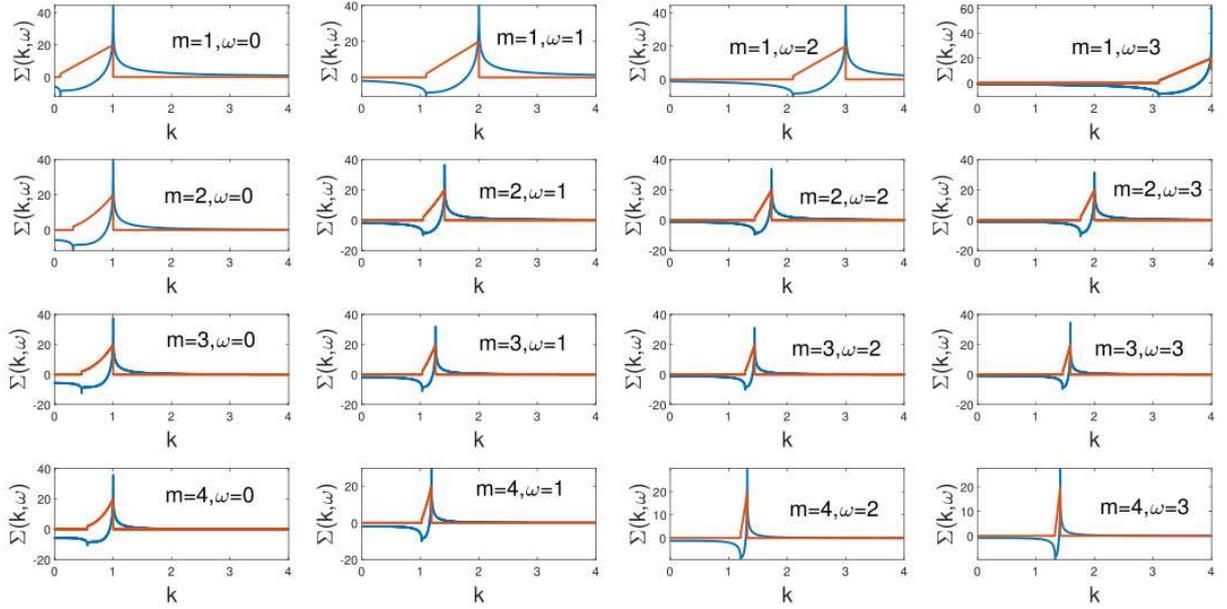}
\caption{Self-energy calculated by the intra-conduction band polarization.
The blue line and red line correspond to the real part and imaginary part, respectively.
}
   \end{center}
\end{figure}

\clearpage
Fig.8
\begin{figure}[!ht]
   \centering
 \centering
   \begin{center}
     \includegraphics*[width=1\linewidth]{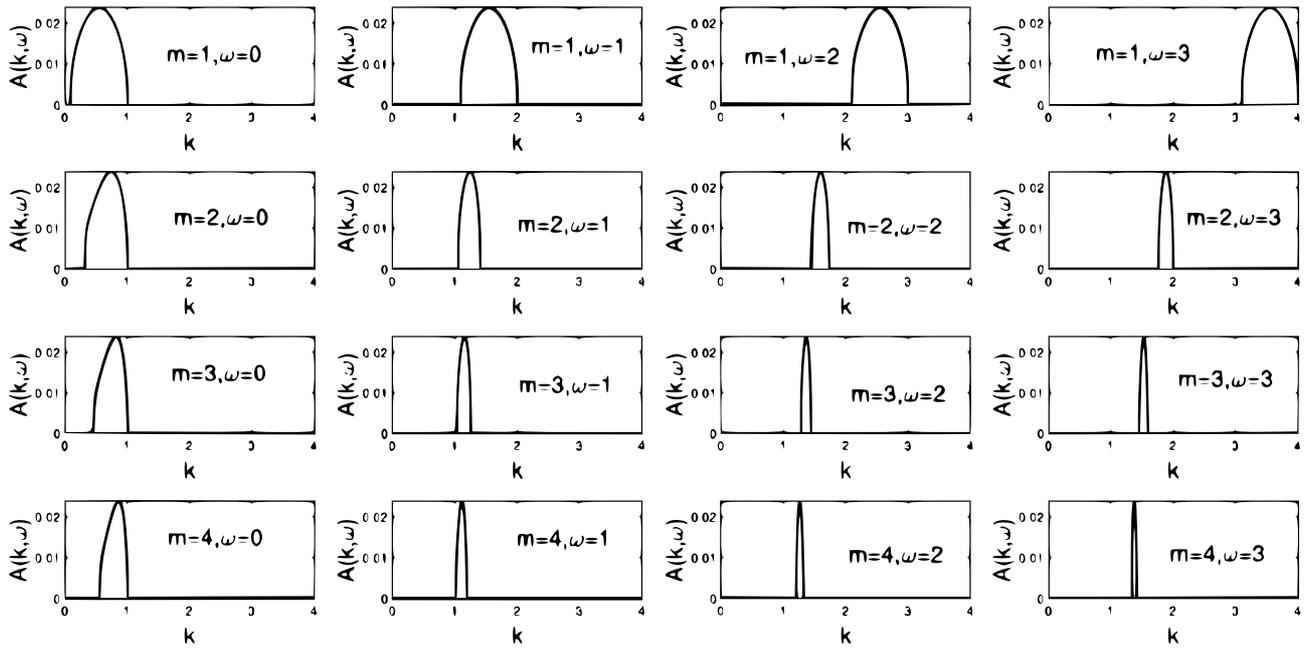}
\caption{Particle spectral function calculated by the intra-conduction band polarization.
}
   \end{center}
\end{figure}

\end{document}